\documentclass[letterpaper,twocolumn,10pt]{article}
\usepackage{usenix-2020-09}
\usepackage{printlen}

\usepackage[textsize=tiny,disable]{todonotes}
\usepackage[toc,acronym,nowarn]{glossaries}
\glsdisablehyper  %
\newacronym{FFI}{FFI}{Foreign Function Interface}
\newacronym{ASAN}{ASan}{AddressSanitizer}
\newacronym{TBI}{TBI}{Top Byte Ignore}
\newacronym{HWASAN}{HWASan}{Hardware-assisted AddressSanitizer}
\newacronym{MLA}{MLA}{Mixed-Language Application}
\newacronym{MIR}{MIR}{Mid-Level Intermediate Representation}
\newacronym{MSB}{MSB}{Most Significant Bit}
\newacronym{LSB}{LSB}{Least Significant Bit}
\newacronym{SSA}{SSA}{Static Single Assignment}
\newacronym{CFG}{CFG}{Control-Flow Graph}
\newacronym{UAF}{UAF}{Use-After-Free}
\newacronym{UAR}{UAR}{Use-After-Return}
\newacronym{CFI}{CFI}{Control-Flow Integrity}
\newacronym{MPK}{MPK}{Memory Protection Keys}
\newacronym{GEP}{GEP}{GetElementPointer}

\usepackage{booktabs}
\usepackage[linesnumbered,ruled,vlined]{algorithm2e}
\usepackage{svg}
\usepackage{subcaption}

\usepackage{inconsolata}

\usepackage{printlen}

\newcommand{\ourApproach}{\mbox{SafeFFI}\xspace}
\newcommand{\SafeFFILib}{\mbox{SafeFFI-Lib}\xspace}

\usepackage{csquotes}
\usepackage{xspace}
\usepackage[inline]{enumitem}
\setlist{noitemsep, topsep=0pt, parsep=0pt, partopsep=0pt} %

\usepackage{listings, listings-rust}
\makeatletter
\lst@CCPutMacro
    \lst@ProcessOther {"2A}{%
      \lst@ttfamily 
         {\raisebox{2pt}{*}}%
         \textasteriskcentered}%
    \@empty\z@\@empty
\makeatother 
\newcommand{\Cpp}{C\texttt{++}\xspace}
\newcommand{\Ccpp}{C/\Cpp}

\definecolor{safecolor}{HTML}{90be6d}
\definecolor{unsafecolor}{HTML}{f94144}
\newcommand{\safe}{\textcolor{black}{\texttt{safe}}\xspace}
\newcommand{\raw}{\textcolor{black}{\texttt{raw}}\xspace}
\newcommand{\noptr}{\textcolor{black}{\texttt{NOPTR}}\xspace}
\newcommand{\unsafe}{\textcolor{black}{\texttt{unsafe}}\xspace}
\newcommand*{\eg}{e.g.,\@\xspace}

\newcommand{\hwasan}{HWASan\xspace}

\newcommand{\FreeBeforeScope}{Free-Before-Scope\xspace}
\newcommand{\FreeDuringScope}{Free-During-Scope\xspace}

\lstdefinestyle{RustCode}{
  language=Rust, style=boxed, basicstyle=\ttfamily\scriptsize,
}

\lstdefinestyle{LLVMCode}{
  language=llvm, style=boxed, basicstyle=\ttfamily\scriptsize,
}

\newlength{\FontHeight}
\newcommand{\MyTikzNode}[1]{\tikz[remember picture]{\coordinate (#1) at (0,0.5\FontHeight);}}

\lstdefinestyle{inlinecode}{
  basicstyle=\ttfamily,
  breaklines=true,
  keywordstyle=\color{blue},
  showstringspaces=false,
  language={[LaTeX]TeX},
  mathescape=true,
  literate={_}{{\textunderscore}}1,
}
\newcommand{\inlinecode}[1]{\mbox{\lstinline[style=inlinecode]{#1}}}

\usepackage{amsthm}

\theoremstyle{definition}

\newtheorem*{theorem*}{Theorem}

\def\Snospace~{\S{}}

\usepackage{xcolor}
\definecolor{blue}{HTML}{5e81ac}
\definecolor{green}{HTML}{a3be8c}
\definecolor{red}{HTML}{bf616a}
\definecolor{darkblue}{HTML}{3b4252}
\definecolor{yellow}{HTML}{ebcb8b}

\usetikzlibrary{shapes,arrows,shadows,calc,graphs,positioning}

\newcommand{\tdob}[1]{\todo[color=orange!60]{\textbf{OB}: #1}}

\newcommand{\tdjh}[1]{\todo[color=green!60]{\textbf{JH}: #1}}
\newcommand{\CommentCameraReady}[1]{\todo[color=blue!60]{\textbf{OB} camera-ready: #1}}
\newcommand{\question}[1]{\todo[color=red]{\scriptsize\textbf{Question}: #1}}

\newcommand{\tdprioB}[1]{\todo[color=orange!50]{\textbf{Prio2}: #1}}
\newcommand{\tdprioC}[1]{\todo[color=orange!20]{\textbf{Prio3}: #1}}

\usepackage[most]{tcolorbox}

\tcbset{
  prunebox/.style={
    colback=yellow!20,
    colframe=yellow!20,
    boxrule=0pt,
    arc=0mm,
    left=0mm,
    right=0mm,
    top=0mm,
    bottom=0mm
  }
}

\newif\ifprunemode

\usepackage{pgfplotstable}
\usepackage{booktabs}
\usepackage{array}
\usepackage{ragged2e}
\newcolumntype{P}[1]{>{\RaggedLeft\hspace{0pt}}p{#1}}

\usepackage{wasysym}
\usepackage{pifont}%
\newcommand{\cmark}{\ding{51}}%
\newcommand{\xmark}{\ding{55}}%

\everymath=\expandafter{\the\everymath\displaystyle}

\newcommand{\HwasanCtOhAvg}{1.69}

\newcommand{\SafeFFINoneCtOhAvg}{2.02}

\newcommand{\SafeFFICtOhAvg}{2.64}

\newcommand{\SafeFFICmOhAvg}{1.14}

\newcommand{\RustSanCtOhAvg}{8.83}

\newcommand{\RustSanCtOhMax}{43.03}

\newcommand{\RustSanCmOhAvg}{9.95}

\newcommand{\RustSanCmOhMax}{59.03}

\newcommand{\SafeFFIUnsafeChecksAvg}{7.38}

\newcommand{\SafeFFIUnsafeAndAddedAvg}{10.22}

\newcommand{\SafeFFIFullChecksAvg}{21.43}

\newcommand{\SafeFFIVsRustSanAvg}{7.34}

\newcommand{\ElisionNumSafeFFIBetter}{28}
\newcommand{\ElisionNumRustSanBetter}{2}
\newcommand{\ElisionNumSafeFFIBetterFivePercent}{24}
\newcommand{\ElisionNumRustSanBetterFivePercent}{1}

\newcommand{\HwasanRtOhAvg}{3.22}
\newcommand{\HwasanRtOhMedian}{3.03}

\newcommand{\SafeFFINoneRtOhAvg}{2.12}
\newcommand{\SafeFFINoneRtOhMedian}{2.06}

\newcommand{\SafeFFIRtOhAvg}{2.51}
\newcommand{\SafeFFIRtOhMedian}{2.24}

\newcommand{\SafeFFINoneNoImprov}{1}

\newcommand{\SafeFFINoImprov}{4}

\newcommand{\NumRtBenchmarks}{20}

\usepackage{authblk}
\begin{document}

\title{\Large \bf \ourApproach: Efficient Sanitization at the Boundary Between Safe and Unsafe Code in Rust and Mixed-Language Applications}

\date{}

\author[1]{Oliver Braunsdorf}
\author[1]{Tim Lange}
\author[2]{Konrad Hohentanner}
\author[2]{Julian Horsch}
\author[1]{Johannes Kinder}
\affil[1]{LMU Munich, Germany}
\affil[2]{Fraunhofer AISEC, Germany}

\maketitle

\begin{abstract}

  Unsafe Rust code is necessary for interoperability with C/C++ libraries and implementing low-level data structures, but it can cause memory safety violations in otherwise memory-safe Rust programs.
  Sanitizers can catch such memory errors at runtime, but introduce many unnecessary checks even for memory accesses guaranteed safe by the Rust type system.
  We introduce SafeFFI, a system for optimizing memory safety instrumentation in Rust binaries such that checks occur at the boundary between unsafe and safe code, handing over the enforcement of memory safety from the sanitizer to the Rust type system.
  Unlike previous approaches, our design avoids expensive whole-program analysis and adds much less compile-time overhead ($\SafeFFICtOhAvg \times$ compared to over $\RustSanCtOhAvg \times$).
  On a collection of popular Rust crates and known vulnerable Rust code, SafeFFI achieves superior performance compared to state-of-the-art systems, reducing sanitizer checks by up to 98\%, while maintaining correctness and flagging all spatial and temporal memory safety violations.
\end{abstract}

\section{Introduction}

Memory corruptions caused by unsafe programming languages such as C and C++ remain a major source of critical software vulnerabilities. Memory bugs regularly take top spots in the lists of most dangerous~\cite{CWETop25} and known exploited weaknesses~\cite{CWETop10KEV}, and studies by Google~\cite{chromium-memsafety,projectzero-zerodays} and Microsoft~\cite{miller2019trends} indicate that around 70\% of their severe bugs are caused by memory unsafety.

In recent years, the Rust programming language has gained traction as a safe-by-construction solution for newly developed software.
Securing existing \Ccpp programs by rewriting them in safe languages requires substantial development effort, however. Despite recent efforts to translate legacy code bases to Rust~\cite{sactor,c2rustbench}, we can expect \Ccpp code to be relied on for the foreseeable future.
Thus, in practice, Rust applications may actually be \emph{mixed language applications} (MLAs), where the Rust application is linked against \Ccpp code---or vice versa---via a foreign function interface (FFI). In this situation, the safety guarantees of Rust may be compromised by unsafe operations in foreign code.

But even pure Rust code may contain \emph{unsafe code}, marked via the \unsafe keyword, allowed to violate the strict typing rules for implementing efficient algorithms and data structures or hardware interactions. 
Thus, be it from foreign functions or explicitly marked unsafe code regions, Rust bears the risk of memory-safety violations originating in unsafe code that potentially affect the whole code base, even safe code~\cite{ERASanEfficientRust2024}.

Both \Ccpp and unsafe Rust code can be protected from exploitation by applying memory-safety \emph{sanitizers}~\cite{SLR+19,vintila-mset2025-sos} to the whole code base, which transparently introduce run-time checks for memory operations. 
While sanitizers such as \gls{ASAN}~\cite{asan} and \gls{HWASAN}~\cite{hwasan} do not require any source code changes and can achieve thorough memory safety guarantees~\cite{vintila-mset2025-sos}, they introduce a significant run-time overhead by inserting checks for \emph{every} pointer dereference in the code. Many of these checks are unnecessary, however: a pointer dereference in Rust is guaranteed safe as long as the pointer is safe and cannot be affected by unsafe code.\tdprioC{OB: last sentence imprecise}
Previous solutions for selective sanitization of Rust code~\cite{RustSanRetrofittingAddressSanitizer2024,ERASanEfficientRust2024} use whole-program static points-to analysis to classify pointers as provably safe or potentially unsafe and then elide checks for an under-approximation of safe pointers. While theoretically sound, this approach incurs significant compile-time overhead and misses optimization opportunities. In contrast, our approach relies on efficient local reasoning about pointer types and their safety guarantees.

In this paper, we present \ourApproach, a new approach to reduce the overhead of memory-safety sanitizers in Rust applications and MLAs consisting of C and Rust code.
\ourApproach utilizes the fact that Rust's strong type system enforces guarantees for pointer types such as reference types (\inlinecode{&T}) and box types (\inlinecode{Box<T>}), while raw pointers (\inlinecode{*const T}) are unchecked.

A key insight in our work is that the cast from a raw pointer type to a safe pointer type forms the boundary between sanitizer-enforced memory safety and type-system-enforced memory safety. 
\ourApproach hoists and bundles checks into a single dynamically-checked precondition, which is propagated statically through the type system. This way, we 
free the memory sanitizer from checking \textit{every} pointer dereference, including from checking dereferences of potentially unsafe pointers.\todo{revise}
Therefore, for most patterns of Rust programs, we can elide more sanitizer checks than previous approaches, leading to better run-time performance.

Because our check placement enforces memory-safety across function calls, we only require local reasoning to ensure memory-safety across the whole application. Hence, we avoid expensive whole-program static analysis leading to better compile-time performance and enabling \ourApproach to scale well on large software projects.
In summary, we make the following contributions:
\begin{itemize}
  \item A novel concept and algorithm for combining sanitizers for unsafe languages and type information from strongly-typed languages to enforce memory safety in mixed code. Our algorithm avoids expensive whole-program static analysis and exposes bugs early by placing checks at the location where the type system expects safety guarantees to hold~(\autoref{sec:check-placement}).
  \item An architecture based on a modified Rust compiler that allows for analyses across multiple intermediate representations of Rust and LLVM to implement the concept, including an efficient algorithm for finding potentially deallocating functions~(\autoref{sec:architecture}).
  \item A systematic evaluation using LLVM's widely-used \gls{HWASAN} on popular Rust crates, known vulnerable Rust code, and a set of microbenchmarks. Compared to the state of the art, we demonstrate that \ourApproach significantly reduces the compile time ($\SafeFFICtOhAvg \times$ compared to over $\RustSanCtOhAvg \times$), increases the rate of elided checks (up to 98\%), and consequently reduces the run-time overhead (from $\HwasanRtOhAvg\times$ to $\SafeFFIRtOhAvg\times$ on average)~(\autoref{sec:evaluation}).
\end{itemize}

\section{Background}
\label{sec:background}

We now introduce the necessary background on memory safety in C/C++ and existing methods for runtime sanitization (\autoref{sec:background_c}), followed by revisiting the memory safety guarantees in Rust and the impact of unsafe code (\autoref{sec:background_Rust-memory-safety}).

\begin{figure*}
\centering
\includegraphics[keepaspectratio,width=0.9\linewidth]{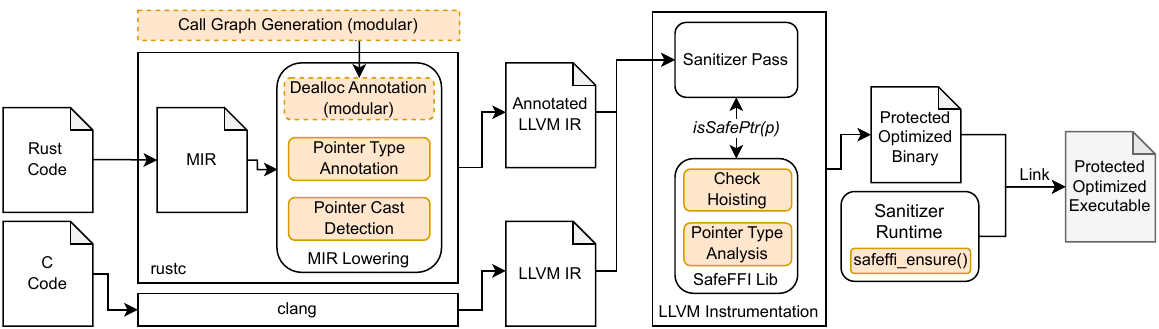}
\caption{\ourApproach Architecture Overview. Rust and \Ccpp are compiled to LLVM IR, a common intermediate representation for sanitizer instrumentation. The orange boxes depict our extensions to rustc and LLVM in this compilation workflow.}
\label{fig:architecture-overview}
\end{figure*}
\CommentCameraReady{JH Fig \ref{fig:architecture-overview}: Just naming the box with the SafeFFI lib "LLVM" seems a bit unspecific, since this is all more or less LLVM; OB: dealloc annotation is optional, call graph generation is modular}

\subsection{Memory Safety and C/C++}
\label{sec:background_c}

C and C++ remain widely used in security-critical software but inherently lack memory safety.
Memory safety violations are classified into \emph{spatial} (e.g., buffer overflows) and \emph{temporal} (e.g., use-after-free) errors.
To catch memory safety bugs, sanitizers~\cite{SLR+19,vintila-mset2025-sos} typically instrument code at compile time with additional check logic to detect violations at runtime.
Furthermore, sanitizers often use symbol interposition to redirect function calls to instrumented versions of a library function, e.g., for malloc, free, or memcpy. This technique is called \emph{Interception} and is especially helpful for sanitizing 3rd-party libraries without recompiling.
Sanitizers use different types of metadata to track memory state and detect violations, typically categorized as being either object-based~\cite{CAMP,chen2023mtsan,HZH23,PACMem,hwasan,asan} or pointer-based~\cite{NZMZ09,OBZH24,kroes2018delta,Intel-PC}.
Object-based metadata, such as redzones~\cite{asan} or memory tags~\cite{hwasan}, is associated with memory allocations and mark memory regions as valid or invalid.
In contrast, pointer-based metadata augments pointers with additional information, such as bounds\cite{NZMZ09,OBZH24} and references to temporal metadata~\cite{NZMZ10}.
Since they are associated with pointers, they can track the validity of memory accesses based on the pointer's state, \eg enabling also sub-object bounds tracking~\cite{OBZH24}.

Two widely used memory-safety sanitizers are \gls{ASAN} \cite{asan} and \gls{HWASAN} \cite{hwasan}, available in LLVM and GCC.
ASan instruments code to place red zones in shadow memory, catching out-of-bounds and use-after-free errors; it also poisons freed memory and delays reuse, improving detection at the cost of higher memory overhead.
HWASan reduces overhead by using tagged pointers and memory tags (typically 16:1 granularity).
Each block carries a tag in shadow memory, and the upper pointer bits hold a matching tag; a mismatch signals a violation.
Retagging freed memory improves temporal detection, while spatial detection remains probabilistic and depends on tag size.
On ARM, \gls{HWASAN} commonly leverages top-byte-ignore (TBI) to store tags in the pointer's top byte, improving compatibility with non-instrumented code and avoiding address-translation issues.

\subsection{Memory Safety and Rust}
\label{sec:background_Rust-memory-safety}

Rust is a modern systems programming language that aims to eliminate memory safety bugs through a strong static type system. 
This type system is built on the principles of ownership, borrowing, and lifetimes~\cite{RustBookOwnershipChapter} which guarantee spatial and temporal memory safety for safe pointer-like types.
\emph{Spatial safety} in Rust is enforced either at run-time or at compile-time.
For statically-sized memory objects, the size of the memory object is known at compile-time. 
The Rust compiler tracks the object's type and its corresponding size for every safe pointer derived from the memory object.
Thus, it can statically verify that a pointer dereference is within the bounds of the pointed-to memory object.
For dynamically-sized types, the compiler generates runtime bounds checks when indexing or dereferencing through a safe pointer.
\emph{Temporal safety} is guaranteed by the type system which enforces that each value is owned by exactly one variable.
The compiler takes care of generating code for the deallocation of the value at the location where this owning variable goes out of scope which prevents double-free errors.
Lifetimes tracked statically by the compiler ensure that references to a value do not outlive the owner, hence preventing \gls{UAF} errors.

\paragraph{Safe Pointer-Like Types.}
\label{sec:bg:safe-pointer-likes}  
The Rust language provides a set of \emph{safe pointer-like} types that are subject to the borrow checker and lifetime analysis. These include:
\begin{itemize}
  \item \texttt{\&T} – shared, immutable reference
  \item \texttt{\&mut T} – exclusive, mutable reference
  \item \texttt{Box<T>} – owning pointer to heap-allocated data
  \item \texttt{[T; N]} – statically sized array
  \item \texttt{\&[T]}, \texttt{\&str} – dynamically sized slice/string
  \item \texttt{\&dyn Trait} – trait object reference, a pointer to an object whose type is only known at runtime
  \item Function pointers and closures
\end{itemize}
In the remainder of the paper, we will usually refer to these types summarily as \safe pointers, in contrast to \raw pointers.

For pointers of these types that originate in \safe Rust code, the Rust compiler ensures that they are non-null, properly aligned, and dereferenceable for the size of their pointee-type~\cite{RustReferenceSafety}.
When \safe pointers are derived from \raw pointers, which may happen in \unsafe code, Rust requires the developer to guarantee those same memory safety properties.

\paragraph{Raw Pointers and Unsafe Code.}
Rust also provides another primitive pointer type, called the \raw pointer: \texttt{*const T} and \texttt{*mut T}.
The \raw pointer type is not subject to borrow checking or lifetime tracking. 
It does not provide any memory safety guarantees, thus it behaves exactly like a pointer in the \Ccpp and also has the same layout.
Raw pointers may only be dereferenced inside functions or code blocks that are marked with the \texttt{unsafe} keyword in Rust.%
Besides dereferencing raw pointers, \unsafe Rust code also enables the following operations which can undermine Rust's safety guarantees:
\begin{enumerate*}
  \item calling other \texttt{unsafe} functions or foreign functions,
  \item accessing mutable static variables,
  \item accessing \texttt{union} fields,
  \item implementing unsafe traits.
\end{enumerate*}
Those \unsafe operations can undermine Rust's typesystem.
Thus, within \texttt{unsafe} blocks, it is the programmer’s responsibility to uphold Rust’s memory-safety guarantees.
Because programmers can make mistakes in \unsafe code, Rust programs might introduce memory-safety violations despite Rust's strong type system.
For such scenarios, the Rust compiler offers the possibility of using memory-safety sanitizers (see \autoref{sec:background_c}).
However, these sanitizers lack awareness of Rust’s static guarantees and therefore redundantly check even \safe Rust code.
This motivates our approach to combine Rust’s compile-time guarantees with selective run-time checks of sanitizers.

\paragraph{Memory Safety in Mixed-Language Applications.}
\label{sec:background_mla}
\glsreset{MLA}
A typical scenario of using \unsafe code in Rust is the invocation of foreign functions written in other languages like \Ccpp.
Rust provides a feature called \gls{FFI} which allows declaring or exporting a function symbol from/to a foreign library which later is linked in by the linker.%
Many real-world applications combine Rust with existing C or C++ code (or vice versa) this way -- they are called \glspl{MLA}.
Raw pointers are used heavily to exchange data across the \gls{FFI} boundary because they have the same layout in Rust as in \Ccpp.
Bugs in the \gls{FFI} definition or in the \Ccpp code parts of an \gls{MLA} can undermine Rust's guarantees and thus affect the safety of the whole application~\cite{DetectingCrosslanguageMemory2022}.
Even worse, cross-language vulnerabilities can be vehicles for bypassing hardening mechanisms for \Ccpp like \gls{CFI}~\cite{CrossLanguageAttacks2022,ExploitingMixedBinaries2021}.
For \glspl{MLA}, memory-safety sanitizers are applicable to protect the whole application because Rust code and \Ccpp code can both be compiled to LLVM IR code, which sanitizers like \gls{ASAN} and \gls{HWASAN} can instrument.

\section{Overview}

\begin{figure*}
    \centering
    \begin{subfigure}[t]{0.30\textwidth}
        \centering
        \begin{lstlisting}[escapeinside=``, style=RustCode, breaklines=true]
fn foo(p: &i32) {

  let n: i32 = *p;
  let raw1: *mut SomeStruct = 
    unsafe { c_create(n) };
  let safe1: &mut SomeStruct = 
    unsafe { &mut *raw1 }; `\label{fig:overview-example:rust-source:ptr-cast}`

  let a: &mut i32 = &mut safe1.a; `\label{fig:overview-example:rust-source:borrow-a}`
  let b: &mut i32 = &mut safe1.b; `\label{fig:overview-example:rust-source:borrow-b}`
  do_some(*a, *b); `\label{fig:overview-example:rust-source:deref-dosome}`



  loop {
    let c: &i32 = derive(a);
    let d: &i32 = derive(b);
    if *c + *d == 85 {
        break;
    }
  }
}
        \end{lstlisting}
        \caption{Rust Code}
        \label{fig:overview-example:rust-source}
    \end{subfigure}
    \begin{subfigure}[t]{0.33\textwidth}
        \centering
        \begin{lstlisting}[escapeinside=``, style=LLVMCode, breaklines=true]
define @foo(ptr %p) {
start:
 sanitizer_check(%p) `\MyTikzNode{ParamCheck}`
 %n = load i32, ptr %p
 %raw1 = call ptr @c_create(i32 %n) `\label{fig:overview-example:llvm-sanitized:rawptr-create}`

 %a = GEP ptr %raw1, i64 8 `\label{fig:overview-example:llvm-sanitized:gep-a}`
 %b = GEP ptr %raw1, i64 12 `\label{fig:overview-example:llvm-sanitized:gep-b}`
 sanitizer_check(%a) `\MyTikzNode{Check2}`
 %0 = load i32, ptr %a `\label{fig:overview-example:llvm-sanitized:load-a}`
 sanitizer_check(%b) `\MyTikzNode{Check3}`
 %1 = load i32, ptr %b `\label{fig:overview-example:llvm-sanitized:load-b}`
 call @do_some(i32 %0, i32 %1)
 br label %bb3


bb3:
 %c = call ptr @derive(ptr %0)
 %d = call ptr @derive(ptr %1)
 sanitizer_check(%c) `\MyTikzNode{Check4}`
 %_14 = load i32, ptr %c
 sanitizer_check(%d) `\MyTikzNode{Check5}`
 %_15 = load i32, ptr %d
 %_13 = add i32 %_15, %_14
 %2 = icmp eq i32 %_13, 85
 br i1 %2, label %bb6, label %bb3

bb6:
 ret void
}
        \end{lstlisting}
        \caption{LLVM IR with sanitizer checks}
        \label{fig:overview-example:llvm-sanitized}
    \end{subfigure}
    \begin{subfigure}[t]{0.34\textwidth}
        \centering
        \begin{lstlisting}[escapeinside=``, style=LLVMCode, breaklines=true, columns=fullflexible]
define @foo(ptr %p safePtrArg(4)) {`\label{fig:overview-example:llvm-optimized:safe}`
start: 
 %n = load i32, ptr %p
 %raw1 = call @c_create(i32 %n) `\textbf{!raw}`
 `\MyTikzNode{SafeFFICheck}`%raw1_safe = call safeffi_ensure(ptr raw1, i64 16) `\textbf{!safe}` `\label{fig:overview-example:llvm-optimized:safeffi_ensure}` 
 %a = GEP ptr %raw1_safe, i64 8 `\label{fig:overview-example:llvm-optimized:gep-a}`
 %b = GEP ptr %raw1_safe, i64 12 `\label{fig:overview-example:llvm-optimized:gep-b}`
 %0 = load i32, ptr %a
 %1 = load i32, ptr %b
 call @do_some(i32 %0, i32 %1)
 br label %bb3




bb3:
 %c = call ptr @derive(ptr %0) `\textbf{!safe}`
 %d = call ptr @derive(ptr %1) `\textbf{!safe}`
 %_14 = load i32, ptr %c
 %_15 = load i32, ptr %d
 %_13 = add i32 %_15, %_14
 %2 = icmp eq i32 %_13, 85
 br i1 %2, label %bb6, label %bb3




bb6:
  ret void
}
        \end{lstlisting}
        \caption{Modified LLVM IR with sanitizer checks being hoisted by \ourApproach}
        \label{fig:overview-example:llvm-optimized}
    \end{subfigure}

    \begin{tikzpicture}[remember picture,overlay]]
      % Define a TikZ style for drawing
      \tikzset{
          myarrow/.style={->, opacity=0.6, thick, blue}
      }
      \draw[myarrow] (Check2) to[out=30, in=180]  (SafeFFICheck);
      \draw[myarrow] (Check3) to[out=0, in=180]   (SafeFFICheck);
      %\draw[myarrow] (Check4) to[out=30, in=180]  (SafeFFICheck);
      %\draw[myarrow] (Check5) to[out=30, in=180]  (SafeFFICheck);
      \draw[-,thick, orange, opacity=0.6] (ParamCheck) to[out=0, in=180]  ($(ParamCheck) + (2,0.5)$);
      \node at ($(ParamCheck) + (2.05,0.5)$) {\textsf{{\textcolor{orange}{\transparent{0.6}x}}}};
      \draw[-,thick, orange, opacity=0.6] (Check4) to[out=0, in=180]  ($(Check4) + (2,1.0)$);
      \draw[-,thick, orange, opacity=0.6] (Check5) to[out=0, in=180]  ($(Check4) + (2,1.0)$);
      \node at ($(Check4) + (2.05,1.0)$) {\textsf{{\textcolor{orange}{\transparent{0.6}x}}}};
    \end{tikzpicture}

    \caption{A simplified example that shows which sanitizer checks can be elided using our \ourApproach optimization}
    \label{fig:overview-example}
\end{figure*}

\autoref{fig:architecture-overview} shows an overview of \ourApproach's architecture. It is implemented as an extension of the Rust compiler pipeline.
We extend the Rust compiler by an analysis of Rust's \gls{MIR} to determine the types of all pointers, locate pointer cast locations, and generate corresponding annotations for the LLVM IR code that is lowered from the \gls{MIR}.
We extend LLVM by our new \SafeFFILib which consumes the pointer type annotation, conducts a pointer type analysis on LLVM IR level, inserts additional sanitizer checks at the boundary between safe and unsafe pointer types, and provides a simple API for existing sanitizers to elide checks for provably safe pointer types.
A more detailed description of the architecture is given in \autoref{sec:architecture}.
\CommentCameraReady{KH: in der Grafik ist nur C Code, wenn auch C++ Code geht würde ich C\/C++ machen. Grafik ist ohne Schattierung vlt übersichtlicher}

\paragraph{Example.}
We intuitively illustrate the effect of \ourApproach 
using the example in \autoref{fig:overview-example}, which contains 
Rust source code of a function \inlinecode{foo}, %
the corresponding LLVM code including sanitizer checks, and the same LLVM code after optimization with \ourApproach.
\CommentCameraReady{KH: die wichigen Stellen (sanitizer\_check, etc) koennte man noch highlighten}
In its normal operation, the sanitizer inserts \inlinecode{sanitizer_check()} for the pointer operands in every LLVM \inlinecode{load} or \inlinecode{store} instructions.
By using information from Rust's typesystem, \ourApproach can deduce that pointers \inlinecode{a}, \inlinecode{b}, \inlinecode{c}, and \inlinecode{d} are all derived from a Rust reference \inlinecode{safe1}.
The Rust type system is guaranteeing that all derived references only ever access memory within the bounds and lifetime of the object they are derived from.
Therefore, if we can ensure that the requirements of Rust's type system (see \autoref{sec:bg:safe-pointer-likes}) are met for the creation of the original safe pointer-like, then all subsequent sanitizer checks are obsolete and we can elide them.

In source line \ref{fig:overview-example:rust-source:ptr-cast}
of \autoref{fig:overview-example:rust-source}, 
the \safe pointer \inlinecode{safe1: &mut SomeStruct} (a mutable reference) is created by casting from the \raw pointer which has been returned from calling the unsafe function \inlinecode{c_create()}.
Later, in lines \ref{fig:overview-example:rust-source:borrow-a} and \ref{fig:overview-example:rust-source:borrow-b}, two more \safe pointers \inlinecode{a} and \inlinecode{b} are derived from \inlinecode{safe1} by taking the respective addresses of the fields defined in \inlinecode{SomeStruct}.
Both are then dereferenced in line \ref{fig:overview-example:rust-source:deref-dosome} to pass their pointee values to the function \inlinecode{do_some()}.
Those dereference operations correspond to the \inlinecode{load} instructions in lines \ref{fig:overview-example:llvm-sanitized:load-a} and \ref{fig:overview-example:llvm-sanitized:load-b} of \autoref{fig:overview-example:llvm-sanitized}.
One can see how the sanitizer inserts \inlinecode{sanitizer_check()} calls before each \inlinecode{load} instruction.
Because Rust's static type system guarantees that \inlinecode{safe1} and its derived pointers \inlinecode{a} and \inlinecode{b} only ever access the memory object within the bounds of the \inlinecode{SomeStruct} object, those sanitizer checks can be elided.
However, to be able to rely on the type system's soundness, we have to ensure that its requirements are upheld when casting \inlinecode{raw1} to \inlinecode{safe1} by inserting a sanitizer check.
This is depicted by the blue arrows: \ourApproach removes the checks before the load instructions and instead inserts the \inlinecode{safeffi_ensure()} call in line \ref{fig:overview-example:llvm-optimized:safeffi_ensure} of \autoref{fig:overview-example:llvm-optimized}.

The \inlinecode{safeffi_ensure()} function takes the \inlinecode{raw1} pointer as input as well as the size of the \inlinecode{SomeStruct} type (in this case \inlinecode{16}) and internally uses a sanitizer check to validate that the memory object behind \inlinecode{raw1} is still allocated, has at least the\CommentCameraReady{is well-aligned?} expected size, and that the pointer has the correct provenance for this object.
If the \inlinecode{safeffi_ensure()} check fails, we immediately abort the program, pointing the developer directly to the cast location. 
This is a specific advantage of \ourApproach: 
it can reveal a misuse of \unsafe code or \gls{FFI} code much earlier compared to normal sanitizer operation. Regular sanitizers only fail at the dereference location which might happen much later, e.g., in a subsequent function call, making the error harder to debug.

If the \inlinecode{safeffi_ensure()} succeeds then it returns a new LLVM value (\inlinecode{\%raw1_safe} in the example) representing the casted \safe version. %
Usually, the Rust compiler would only generate one LLVM variable to represent both the \raw and the \safe pointer as an optimization, because they have the same machine layout and carry the same value.
\ourApproach allows us instead to differentiate \raw and \safe pointers on LLVM IR level and to elide sanitizer checks accordingly.

In the following section, we give a detailed explanation of our approach to determine in which cases sanitizer checks can be elided and when the boundary between \raw and \safe pointers needs validation at run-time by inserting additional sanitizer checks.

\section{Typesystem-Guided Sanitizer Checks}
\label{sec:check-placement}

Now we explain our concept in detail. We begin by motivating pointer casts as logical boundaries for memory safety enforcement (\autoref{sec:concept_pointer-casts-boundary}). We introduce our method for using function-local type information to hoist checks to casts while eliding sanitizer checks for \safe pointers (\autoref{sec:local-hosting}) and discuss additional measures required for full temporal memory safety (\autoref{sec:heap-temporal}).
Finally, we show how to leverage the Rust type system to maintain memory-safety across function calls without expensive whole-program analysis (\autoref{sec:interprocedural-invariant}).
Note that this design includes full support for the presence of unsafe foreign code in mixed-language scenarios.

\subsection{Memory Safety Enforcement Boundaries}
\label{sec:concept_pointer-casts-boundary}
Incorrect memory access through raw pointers can cause memory bugs even in safe Rust~\cite{ERASanEfficientRust2024}.
Raw pointer dereferences are dangerous because Rust does not enforce any guarantees for \raw pointers.
For \safe pointers, however, the Rust compiler 
provides the following memory-safety guarantees
(see \autoref{sec:background_Rust-memory-safety}):
\begin{itemize}
  \item \emph{Spatial safety}: every read or write through such a pointer is guaranteed to only access memory inside the bounds of its pointee type. 
  \item \emph{Temporal safety}: no \safe pointer can exist outside the lifetime of its original pointed-to memory object.
\end{itemize}
There are multiple ways to create and derive \safe pointers. All of them are checked by the Rust compiler, \emph{except} for the cast from \raw pointer to \safe pointer.
Casts from \raw to \safe pointers can only happen in \unsafe Rust code, where the Rust compiler does not check that the \raw pointer adheres to the requirements of the \safe pointer type.
Thus, a cast of an invalid \raw pointer can undermine the memory safety guarantees of Rust's type system.

The key idea behind \ourApproach is that using sanitizer checks, we can dynamically guarantee validity for a \raw pointer at the time of the cast and then continue to rely on the guarantees statically enforced by the Rust compiler for the remaining lifetime of the \safe pointer after the cast.
Hence, we consider these cast operations to form the boundary between memory safety enforcement by the sanitizer and the Rust compiler.
Casts from \raw to \safe pointers are the central point of focus for \ourApproach, and in the following we show how to hoist the sanitizer checks for \safe pointers by protecting this boundary with a specific sanitizer check.

\subsection{Local Type Analysis for Check Hoisting}
\label{sec:local-hosting}
The basic idea of \ourApproach's optimizations is to use function-local pointer type information provided by annotation of local variables, globals, and arguments to reduce the number of sanitizer checks for \safe pointers.
Usually, the sanitizer inserts checks at every instruction that dereferences a pointer.
For \raw pointers, we just keep all those checks in place. 
For \safe pointers, we hoist the checks from the dereference location up to the beginning of their scope, where they are created.
Then, based on the rules enforced by Rust's type system, we can safely elide all subsequent checks.

We differentiate the following cases of how a \safe pointer can be created in the current function's scope:
\begin{enumerate}[label=(\alph*)]
\item\label{safeptr-create-stack} Allocating a new stack object. On LLVM-IR level this corresponds to an \inlinecode{alloca} instruction which returns a pointer variable.
\item\label{safeptr-create-stack-derive} Deriving a reference from a local stack object or from another safe Rust reference.
\item\label{safeptr-create-cast} Casting a \raw pointer to a reference via a Reborrow operation (\inlinecode{safe_ptr: &T = &*raw_ptr}) or to a Box via \inlinecode{Box<T>::from_raw(raw_ptr)}. Although its syntax looks like a dereference-and-take-address operation, a Reborrow just creates a reference that points to the same memory object as pointer \inlinecode{a1}, without dereferencing.
\item\label{safeptr-create-load} Loading a \safe pointer from memory, e.g., as a field of another object that resides in memory.
\end{enumerate}
In cases \ref{safeptr-create-stack} and \ref{safeptr-create-stack-derive}, Rust takes care of memory management through its lifetime and borrowing mechanics and no explicit \raw pointers are involved, so we can elide all checks.
In case \ref{safeptr-create-cast}, we have to ensure the validity of the \raw pointer before we can safely cast it to a \safe pointer and rely on the Rust compiler for memory safety.
To check the validity of \safe pointers casted from \raw pointers, \ourApproach emits a call to \inlinecode{safeffi_ensure()}, a custom dynamic check function using sanitizer metadata to determine whether the object pointed to by the raw pointer is still alive and is at least of the size of the pointee type \inlinecode{T}\CommentCameraReady{and well-aligned}.
For case \ref{safeptr-create-load}, we also need to insert a sanitizer check. 
Since \unsafe Rust or foreign code can arbitrarily modify memory contents, pointers stored on either heap or stack might be corrupted, so we must check their validity when they are loaded into the current function scope.

As a result, we elide sanitizer checks at dereference locations for \safe pointers; if the pointer is not safe by construction, we insert a sanitizer check at the creation site of the \safe pointer, effectively combining and hosting the checks. %
This is illustrated by the arrows in \autoref{fig:overview-example}.
The run-time benefit of removing the sanitizer checks is most pronounced when checks can be hoisted out of loops.
\CommentCameraReady{This would benefit from the new GCD example}

\paragraph{Spatial Safety.}
The inserted sanitizer check enforces the size requirement of the \safe pointer type at the cast site.
Once the \safe pointer is created, we can rely on the Rust type system to ensure that all subsequent accesses through the \safe pointer in the current function are within the bounds of the type, as explained above.
Thus, \ourApproach ensures spatial safety for the whole scope of the \safe pointer (to the extent that the underlying sanitizer guarantees it).

\paragraph{Temporal Safety.}
\label{sec:concept:temporal-safety}
For reasoning about temporal safety, we distinguish \emph{\FreeBeforeScope} vulnerabilities from \emph{\FreeDuringScope} vulnerabilities.
So far, we combined the Rust type system with dynamic sanitizer checks\CommentCameraReady{at cast and load sites} to ensure that any \safe pointer points to a memory object that is alive \textit{at the start of the pointer's scope}.
Thus, \ourApproach always reliably detects all temporal vulnerabilities where the memory object is deallocated \emph{before} the start of the pointer's scope, i.e., \FreeBeforeScope vulnerabilities.
If the memory object is deallocated \emph{during} the \safe pointer's scope, e.g., through an alias pointer, this can cause a \FreeDuringScope vulnerability.
To also catch these, \ourApproach provides the option for further checks, as 
we explain in the next section.

\subsection{Catching \FreeDuringScope Violations}
\label{sec:heap-temporal}
For \FreeDuringScope violations, we need to check that the \safe pointer remains valid until the end of its scope.
\ourApproach provides the option to detect \FreeDuringScope violations which allows developers to trade safety for run-time and compile-time performance.

In Rust, memory may be deallocated either by popping a stack frame at the end of a function scope or by calling a heap deallocation function. %
We reason about a \safe pointer's local scope within the current function---we will extend our reasoning across function calls in \autoref{sec:interprocedural-invariant}.
Because the \safe pointer's scope is limited to the current function, this stack frame is guaranteed to not be deallocated within the entire scope.
Thus, only heap deallocations remain for potential \FreeDuringScope violations.

Heap deallocation requires calling \inlinecode{__rust_dealloc()} or \inlinecode{libc::free()}.
Hence, a \safe pointer can only become dangling within its scope in the current function if there is a call to one of those deallocation functions in between.
To detect \FreeDuringScope violations, \ourApproach inserts an additional \inlinecode{safeffi_ensure()} check for every dereference of the \safe pointer that is reachable from a call to a potential deallocation function.
\autoref{algo:reinsert-heap-checks} for inserting those heap checks can be implemented efficiently within LLVM and is linear in the number of instructions in the analyzed function.
\begin{algorithm}[h]
	\caption{Insert additional heap checks for \FreeDuringScope violations.}
	\label{algo:reinsert-heap-checks}
	\KwIn{A Rust function \textit{F} and the set \textit{DeallocFns} of functions that may (transitively) lead to heap deallocation}
	\ForEach{memInst $\in$ MemoryInstructions(F)}{
    \ForEach{callInst $\in$ memInst.operands()}{
      \If{hasSafePointerOperand(memInst)}{
        \If{callInst.callee() $\in$ DeallocFns}{
          \If{IsReachable(memInst, callInst)}{
              InsertCheckAt(memInst, callInst)\;
          }
        }
      }
    }
	}
\end{algorithm}
To determine the set \textit{DeallocFns} of functions that may transitively lead to a heap deallocation, we develop an efficient and sound analysis for constructing the call-graph, which executes on-the-fly during compilation (see \autoref{sec:architecture:callgraph_based_heap_deallocation} for a detailed description).

In single-threaded applications, inserting additional \inlinecode{safeffi_ensure()} checks guarantees that each \safe pointer remains valid throughout its scope, allowing \ourApproach to reliably detect \FreeDuringScope violations. 
For \FreeDuringScope violations, check hoisting is not thread-safe, however. A deallocation could occur concurrently between the check and the subsequent dereference, which could allow a \FreeDuringScope violation to go undetected. 
Nevertheless, \ourApproach remains fully compatible with multi-threaded programs and is guaranteed to detect spatial and \FreeBeforeScope violations to the extent that the underlying sanitizer does.

\subsection{Interprocedural Memory Safety}
\label{sec:interprocedural-invariant}
Intraprocedurally, \ourApproach 
establishes a safety invariant for every \safe pointer created in each function:
\emph{each \safe pointer created in the current function is guaranteed to be safe to dereference for its whole scope in the current function.}
To guarantee whole-program memory safety, we also need to ensure validity of \safe pointers passed across function calls. %

\paragraph{Pointer Arguments.}
\ourApproach generates annotations for the function signature during lowering from MIR to LLVM IR, emitting attributes for pointer parameters (e.g., see the \inlinecode{safePtrArg} attribute in line \ref{fig:overview-example:llvm-optimized:safe} of \autoref{fig:overview-example:llvm-optimized} for the parameter \inlinecode{p: &i32}).
When the currently analyzed Rust function is called from another Rust function, the type system guarantees that the argument types in the call and the function signature match, so functions with \safe pointer arguments will only ever receive \safe pointer values from the caller.

For Rust functions with external visibility, which can be called by foreign code, invalid data could be passed to a parameter of a \safe pointer type, because linking foreign code requires only ABI compatibility and may ignore types.
Therefore, \ourApproach inserts an additional check in the prologue of the callee to ensure that the \safe pointer is indeed valid.
Because of our established memory safety invariant, we guarantee that \safe pointers are indeed safe to dereference for the remainder of the scope of the caller function, which means they are also valid for the whole scope of the callee.
Hence, we can elide all sanitizer checks for \safe parameters.

\paragraph{Pointer Return Values.} Similar to the previous case, we can rely on the Rust type system to guarantee that the return value of a function call is of the correct type. Because of our invariant, we can guarantee that a \safe pointer returned from a function call is valid until the end of its scope in the callee.
And if the pointer is valid at the end of the callee, then it is also valid at the return in the caller---with one exceptional case that needs special handling.

Each return instruction is also the deallocation of the callee's stack frame. 
If the \safe pointer is derived from a \raw pointer pointing to an object on that same stack frame, then this creates a stack \gls{UAR} violation because by the time the \safe pointer is received in the caller function, the pointed-to memory object on the callee's stack frame is already deallocated.
\autoref{fig:example-stack-temporal} shows an example of such a stack temporal violation.
The call to function \inlinecode{derive()} returns a \safe pointer which has been derived from a \raw pointer pointing to an object on the stack frame.
Because the developer did not choose the correct lifetime for the returned pointer in the \inlinecode{derive()} signature and because foreign C code is involved, the Rust compiler has no chance at preventing this \gls{UAR} violation.
To catch such violations, \ourApproach inserts an additional \inlinecode{safeffi_ensure()} check after every function call that returns a \safe pointer.
Thus, we can now guarantee that the \safe pointer is valid for the scope in the caller function.

\begin{figure}
\lstset{emph={__safeffi_ensure}, emphstyle=\itshape}
\begin{lstlisting}[escapeinside=``, style=RustCode, breaklines=true]
// C code
void* c_derive(int* n) {
  int* p = n;
  ...
  return p;
}

// Rust code
fn derive(_: &'a i32) -> &'a i32 {
  let n = 42;
  // p points to n on the stack
  let p: *const i32 = unsafe { c_derive(&n as *const i32) }; 
  __safeffi_ensure(p, sizeof(i32));
  // at this point, *p is still valid, so cast check succeeds
  let p_safe: &i32 = unsafe { &*p };
  return p_safe;
  // n is deallocated here, so the p_safe pointer is now dangling
}

fn foo() {
  ...
  let b: &i32 = derive(a);
  // additional check catches invalid pointer
  __safeffi_ensure(b, sizeof(i32));

}
\end{lstlisting}
\caption{Example of a stack temporal violation that \ourApproach catches by inserting an additional sanitizer check after the pointer returned in the caller function \inlinecode{foo()}.}
\label{fig:example-stack-temporal}
\end{figure}

\section{\ourApproach's Implementation}
\label{sec:architecture}
In this section, we describe the architecture and implementation details of our approach.
We implemented \ourApproach as modifications to the Rust compiler version 1.52.0 and LLVM version 12.
The blue boxes in \autoref{fig:architecture-overview} highlight how \ourApproach integrates into the Rust and LLVM toolchain. 

First, rustc lowers Rust source code to the \glsreset{MIR}\emph{\gls{MIR}} which is where the Rust type system implements its static checks, aka. the \textit{Borrow Checker}.
While MIR is then lowered to LLVM IR \ourApproach attaches pointer type annotations and inserts sanitizer checks for pointer casts. The LLVM IR is processed by standard optimization and sanitizer passes, and finally linked with the sanitizer runtime. In mixed-language applications, C code is compiled to LLVM IR and linked in the same way.
Further details on each component are described in the following subsections.

The goal for our architecture was to make integration with existing sanitizers as easy as possible by requiring only minimal changes to the sanitizer:
\begin{enumerate*}
 \item querying pointer types via \inlinecode{is_safe_pointer(Value* ptr)} in \SafeFFILib, and 
 \item providing a run-time check implementation \inlinecode{safeffi_ensure(void* p, u64 size)}.
\end{enumerate*}
This way, our architecture increases the chance of adoption.

\subsection{Pointer Type Annotation in MIR}
\label{sec:architecture:pointer-type-annotation}
Our goal is to know the type of every pointer in LLVM IR, because for every pointer, the sanitizer can potentially request the type via the \inlinecode{isSafePtr()} API of \SafeFFILib.
\tdprioC{KH: vlt müsste man wo explizit erwähnen dass in der LLVM version pointer noch types haben, und auch ob/wie das Konzept in neueren clang versionen ohne IR pointer types funktionieren würde. OB: "its about safe vs raw types, we do not care about the actual type, so it would work on newer LLVM versions with opaque pointers"}
Thus we have to annotate local variables, function arguments, global variables, and constants.
We implement this by associating LLVM metadata nodes (\inlinecode{MDNode}) with the corresponding generated LLVM instructions and LLVM \inlinecode{Attributes} for function parameter values.

For this we hook into the lowering process from MIR to LLVM IR. 
The lowering process is implemented in rustc via the visitor pattern: the \inlinecode{rustc_codegen_ssa} module visits MIR statements and calls the functions in \inlinecode{rustc_codegen_llvm} (the LLVM backend interface) to generate the corresponding LLVM IR.
The challenge lies in the loose connection between MIR symbols and corresponding LLVM symbols because the lowering of an MIR symbol depends on the target ABI of its type. 
The Rust compiler differentiates between the following ABIs for any MIR type\footnote{There is no official specification of Rust, thus we have to take the Rust compiler's behaviour as reference: \url{https://github.com/rust-lang/rust/blob/1.52.0/compiler/rustc_target/src/abi/mod.rs\#L847}}:
\begin{itemize}
\item \inlinecode{Uninhabited}: a zero-sized type, that actually does not exist in memory. No LLVM code will be generated for it, so there is nothing to annotate.
\item \inlinecode{Scalar}: a type that is represented by a single LLVM value. If this is a pointer type (also called \textit{thin pointer}), then we annotate it accordingly. This is the case for references and arrays, and raw pointers but it is also the case for all ADT\tdprioB{ADT?} types that are represented by a single pointer value in LLVM IR, e.g. the \inlinecode{Box<T>} or any custom arbitrarily nested struct that only has one field and that field is a pointer. A special case of this is Rust's \inlinecode{Union} type. If it has a \inlinecode{Scalar} ABI, then we can only annotate it as \safe if all of its fields are \safe pointers, otherwise the pointer value could be manipulated in an unsafe way through another type representation like an integer or a raw pointer. We implemented a recursive type analysis to detect such cases.
\item \inlinecode{ScalarPair}: a type that is represented by two LLVM scalar values, mostly dynamically-sized types (e.g. slices) which are lowered to \text{fat pointers} which contain a data pointer and a dynamic size value. As we cannot reason about the dynamic size of such types, \ourApproach annotates them as \raw. We leave it to future work to find further optimizations to elide checks for such types. However, in the case of Trait objects (\inlinecode{&dyn Trait}) the second value is a vtable pointer. Because it is lowered to a LLVM pointer, we need to annotate it too. Because the vtable pointer is not user-manipulated but generated and managed by the compiler, we always annotate it as \safe.
\item \inlinecode{Vector}: those are only used for LLVM's SIMD vector types, which are not relevant for our approach. \CommentCameraReady{Because Sanitizers do not support SIMD vector types?}
\item \inlinecode{Aggregate}: a type that is represented by custom LLVM structs. Those types are actually no pointers in MIR, however, if a local variable of an \inlinecode{Aggregate} type is allocated on the stack using a LLVM \inlinecode{alloca} instruction, then a LLVM pointer is created to represent this variable. 
We are annotating them with a \noptr tag and treat them as always safe because they cannot be user-manipulated because they are compiler-generated pointers.
\end{itemize}

The ABI of the type and the calling convention also dictate how a value is passed or returned as function argument in a call.
Fortunately, the Rust compiler already annotates parameters of the function signature if they are safe to dereference using LLVM's \inlinecode{dereferenceable(<size>)} attribute from which we inherit our annotations.

\subsection{Pointer Type Analysis on LLVM IR}
\label{sec:architecture:pointer-type-analysis}
A further challenge is that the ABI of MIR types also controls how values are loaded from memory. For example, accessing a MIR field (\inlinecode{let a = safe1.a}) can generate different sequences of LLVM instructions like \inlinecode{GEP, BITCAST, LOAD}. Since the Rust compiler does not track all LLVM instructions generated for a MIR value, \ourApproach annotates only the final LLVM value representing the loaded MIR operand, lacking pointer type annotations for intermediate LLVM pointer values.
Thus, in \SafeFFILib, we implement an intra-procedural pointer type analysis that forwards the types (\safe, \raw, and \noptr) throughout each function.
This analysis initializes a map of pointer types using the annotations inserted by the Rust compiler for Alloca Instructions, Call/Invoke instructions, function parameters, and global variables.
Then, for every remaining LLVM instruction that produces a pointer value (BITCAST, GEP, INTTOPTR, PHI), \ourApproach forwards the type from the operands of the instruction to the resulting pointer based on the semantics of the instruction. The \glsreset{GEP}\gls{GEP} instruction, which computes a derived pointer by offsetting a base pointer, is handled more carefully: if the base pointer is \raw, the result is always marked as \raw. If the base pointer is \safe, we attempt to statically compute the offset and check whether it remains within the bounds of the original allocation. If so, \ourApproach marks the result as \safe and otherwise conservatively downgrades it to \raw.

As a result, \emph{every} pointer value in LLVM IR is classified as \safe, \raw, or \noptr, enabling the sanitizer to reliably query pointer types via the \inlinecode{isSafePtr()} API.

\subsection{Pointer Cast Detection in MIR}
\label{sec:architecture:pointer-cast-detection}
The goal of this component is to detect casts from \raw pointers to \safe pointers in MIR.
Our definition of \safe pointers distinguishes 3 types of pointers: References, Boxes, and static arrays.
Raw pointers cannot be casted to static arrays (only to references to static arrays)\CommentCameraReady{@Ian, there is nothing to cite here?}, thus, we only have to detect the following two cases.
\begin{enumerate*}
  \item \textit{Casting a raw pointer to a Box}: This always has to happen through a call to \inlinecode{Box::from\_raw(p)} which is an explicit statement in MIR.
  \item \textit{Casting a raw pointer to a reference:} This too is always an explicit statement in MIR, because Rust does not allow implicit coercion in this case~\cite{RustReferenceTypeCoercions}. In MIR this pattern looks like \inlinecode{q = &*p}, where \inlinecode{&} denotes the creation of a reference and \inlinecode{*} denotes a dereference operation. Because the dereferenced \inlinecode{p} can be any kind of Rust pointer, we have to check if \inlinecode{p} actually has the raw pointer type to confirm this is a cast.
\end{enumerate*}

With both cast types, \inlinecode{p} can be a projection, e.g., the access of a field of a (nested) struct (\inlinecode{&*a.b.c}) or an array (\inlinecode{&*x[y][z]}).%
\ourApproach determines whether the inner-most element is a raw pointer by iterating over the MIR projections until we find the type of the accessed element.
\tdob{Everything above is algorithmic and might be moved into section 4}

We implement the pointer cast detection by hooking into the MIR visitor for \inlinecode{Rvalues} (right-hand-side values) of MIR \inlinecode{Assign} statements.
During lowering of an \inlinecode{RValue}, the Rust compiler generates a series of LLVM IR instructions, the last of which producing an LLVM Value that corresponds to the MIR \inlinecode{Rvalue}. 
For each \inlinecode{Rvalue}, Rustc keeps a mapping from the MIR \inlinecode{Rvalue} to the corresponding LLVM value.
This is where \ourApproach inserts a sanitizer check for pointer casts by extending the series of generated LLVM instructions with a call to the \inlinecode{safeffi_ensure()} function.

We show this effect in \autoref{fig:overview-example}.
The \raw pointer \inlinecode{raw1} is lowered by generating a call to \inlinecode{c_create()} in \autoref{fig:overview-example:llvm-sanitized} line \ref{fig:overview-example:llvm-sanitized:rawptr-create} and its returned LLVM value \inlinecode{\%raw1} is mapped for the \inlinecode{Rvalue}.
For lowering the pointer cast \inlinecode{safe1 = unsafe {&mut *raw1}} the Rust compiler usually just maps the \inlinecode{Rvalue} to the same value as the \raw pointer because there is no difference in the LLVM representation of a \raw pointer and a \safe pointer.
\tdprioB{from above: they have the same machine layout - a word-sized ... - and carry the same value. That is why casts normally are a noop-operation on LLVM IR level.}
Thus, in lines \ref{fig:overview-example:llvm-sanitized:gep-a} and \ref{fig:overview-example:llvm-sanitized:gep-b}, the LLVM value \inlinecode{\%raw1} is used to access the fields of the \inlinecode{safe1} reference.
One can see how this is changed by \ourApproach in the \autoref{fig:overview-example:llvm-optimized} lines \ref{fig:overview-example:llvm-optimized:safeffi_ensure}-\ref{fig:overview-example:llvm-optimized:gep-b}.
Our rustc modifications insert the \inlinecode{safeffi_ensure()} call and generate a new LLVM value \inlinecode{\%raw1_safe}. 
\ourApproach then replaces the \inlinecode{RValue} mapping for cast operation, now mapping to \inlinecode{\%raw1_safe}.
Thus, the subsequent accesses of the fields behind the \safe pointer \inlinecode{safe1} now use the LLVM value \inlinecode{\%raw1_safe}.

Generating a new LLVM value for the casted pointer has the advantage that we can now differentiate between the \raw pointer and the \safe pointer on LLVM IR level and separately reason about their safety guarantees at every subsequent location of use in LLVM IR.
Moreover, our experiments showed that implementing cast detection on MIR level is the most reliable way to detect casts and insert sanitizer checks consistently.
Other attempts to detect casts by finding differences in pointer type annotations at LLVM IR level have shown to be unstable because type annotations are transported via LLVM metadata nodes which can get lost or moved around during optimization passes in LLVM.

\subsection{Callgraph-Based Deallocation Checking}
\label{sec:architecture:callgraph_based_heap_deallocation}
\begin{figure}
  \centering
  \includegraphics[width=.9\linewidth, keepaspectratio]{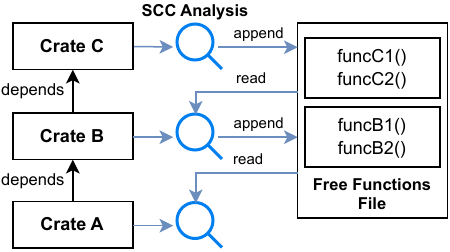}
  \caption{The NoFree SCC Analysis is performed on all dependencies to determine possibly heap-deallocating functions.}
  \label{fig:callgraph_heap_deallocation}
\end{figure}
As mentioned in \autoref{sec:heap-temporal}, detecting \FreeDuringScope temporal vulnerabilities requires checking that a dereferenced pointer has not become dangling since its creation.

To address this, we implemented a call-graph-based heap deallocation analysis in LLVM to determine whether a given function may perform a deallocation.
It analyzes bottom-up over the call graph: functions without call instructions are visited first.
The analysis is implemented as a LLVM SCC (strongly connected components) pass because the call graph might contain cycles.
A function is annotated as \emph{nofree} if it does \emph{not} contain any calls to: (i) known deallocation functions (e.g., \inlinecode{free()} or \inlinecode{__rustc_dealloc()}), (ii) functions without a nofree annotation, or (iii) unknown callees. 
If any function within an SCC cannot proven to be nofree, then the entire SCC treated as potentially deallocating.

Another challenge is that deallocations can happen outside the current compilation unit, e.g., in a Rust dependency or external C library, and thus are not visible to the current compilation process. 
To address this, we serialize the nofree annotations to a persistent file after the analysis has finished for the current compilation unit. 
Subsequent compilation processes read in the serialized annotations and restore them before running the analysis.
This compositional cross-crate analysis is illustrated in \autoref{fig:callgraph_heap_deallocation}.
To support \Ccpp dependencies, we provide build flags for clang to include our analysis.
When C dependencies are dynamically linked or precompiled, we conservatively assume all external C functions may perform deallocations.
bin 
\section{Evaluation}
\label{sec:evaluation}
We implemented \ourApproach on top of Rust nightly-2021-02-22, which corresponds to Rust compiler version 1.52.0 and LLVM version 12.
We integrated \ourApproach into \gls{HWASAN} because it is the sanitizer with the best detection capabilities under the maintained sanitizers\cite{vintila-mset2025-sos}.
Because \gls{HWASAN} uses ARM-specifc hardware features, we ran all experiments with \ourApproach on a MacStudio M2 Ultra with 24 ARMv8 cores and 192 GB RAM running an Ubuntu 20.04 docker container on ASAHI Linux 6.12.0.
To reproduce results of related work as closely as possible and compare to them, we conducted experiments with ERASan~\cite{ERASanEfficientRust2024} and RustSan~\cite{RustSanRetrofittingAddressSanitizer2024} on a x86\_64 system---the architecture on which they were developed---running Ubuntu 24.04. We make all datasets, scripts, and tools used in our evaluation available for reproduction.

In this section, we present our evaluation of \ourApproach answering the following research questions:
\begin{enumerate}[label=\textbf{RQ\arabic*}, align=left]
  \item \label{rq:knownvulns} How does \ourApproach affect the detection capabilities of the sanitizer~(\autoref{sec:correctness})?
  \item \label{rq:elided_checks} How many sanitizer checks can \ourApproach reduce in sanitized programs~(\autoref{sec:eval:effectivness})?
  \item \label{rq:runtime} How much can \ourApproach reduce the run-time of sanitized programs~(\autoref{sec:eval:effectivness})?
  \item \label{rq:artifical_exhaustive} Is \ourApproach's implementation robust in the presence of FFI interactions in \glspl{MLA}~(\autoref{sec:eval:mla})?
  \item \label{rq:compiletime} How much compile-time overhead does \ourApproach incur~(\autoref{sec:compile-time-performance})?
\end{enumerate}

\subsection{Correctness}
\label{sec:correctness}
We answer \ref{rq:knownvulns} by testing \ourApproach on known real-world memory safety vulnerabilities.
We further include RustSan~\cite{RustSanRetrofittingAddressSanitizer2024} and ERASan~\cite{ERASanEfficientRust2024} as related work and \gls{ASAN} as a baseline for those two approaches.
We assembled the dataset of known vulnerabilities by merging and deduplicating the datasets used by ERASan and RustSan. Note that the authors of RustSan have not provided their dataset, thus, we manually reconstructed it based on the RustSec advisories.

\autoref{tab:knownvulns} shows the results of our evaluation on the dataset of known vulnerabilities. \LEFTcircle{} indicates that the vulnerability was detected by the sanitizer, but with a different error message than the baseline. For example, CVE-2021-30457 is both a use-after-free and \emph{later} a double-free. The former is detected by instrumentation and susceptible to check elision while the latter is caught by the sanitizer's interceptor of \inlinecode{free()}. 
Segmentation faults and other signals are classified as not detected (\Circle), as they imply that a memory error was not caught by the sanitizer before leading to the crash. \ourApproach reports some vulnerabilities \emph{earlier} at the root cause instead of at the access location due to the hoisting of checks (cf.~\autoref{sec:local-hosting}), these cases are classified as detected (\CIRCLE{}). %

The results show that \ourApproach catches all vulnerabilities that \gls{HWASAN} detects. 
Due to hoisting checks to the memory safety boundary at the location of pointer casts, \ourApproach is able to report 21 vulnerabilities earlier than \gls{HWASAN}, increasing debuggability, and it detects one vulnerability where every other approach fails.\CommentCameraReady{why?}
Additionally, we have not encountered any false positives during the evaluation of the other research questions as shown in the following sections, which further indicates that \ourApproach does not introduce false positives.

RustSan retains nearly all capabilities of \gls{ASAN}, performing slightly worse than \gls{HWASAN} and \ourApproach in a few cases. In contrast, ERASan performs significantly worse than \ourApproach and RustSan. We investigated further and discovered that ERASan's implementation does not properly annotate all unsafe pointers and therefore removes checks that would have been necess ary to detect the vulnerabilities. Yet, some vulnerabilities are detected because ERASan keeps ASan's interceptors in place, e.g., for \inlinecode{free()} and \inlinecode{memcpy()}. We reached out to the authors of ERASan to debug our findings but have not received a response.\footnote{Github link redacted for review, anonymized screenshot in \autoref{appendix:erasan-issue}}\CommentCameraReady{put issue link}

\begin{table}[th!]
  \caption{Vulnerability Detection}%
  \label{tab:knownvulns}%
  \pgfplotstableread[col sep=comma]{data/correctness.csv}\correctnessstats%
  \pgfplotstablecreatecol[copy column from table={\correctnessstats}{[index] 0}] {ID} {\correctnessstats}
  \scriptsize
  \begin{center}
    \setlength{\tabcolsep}{9pt}
    \vspace*{-5mm}
    \pgfplotstabletypeset[
        columns={ID,INTERCEPT,HWASAN,SafeFFI,ASAN,RustSan,ERASan},
        columns/ID/.style={column type={l}, column type/.add={}{@{\hspace{7mm}}}},
        string type,
        string replace={FULL}{\CIRCLE},
        string replace={FULLSTAR}{\phantom{\textsuperscript{\textdagger}}\CIRCLE{}\textsuperscript{\textdagger}},
        string replace={EMPTY}{\Circle},
        string replace={HALF}{\LEFTcircle},
        string replace={TRUE}{\cmark},
        string replace={FALSE}{\xmark},
        every head row/.style={
          before row=\toprule,
          after row=\midrule,
          typeset cell/.code={
            \ifnum\pgfplotstablecol=\pgfplotstablecols
            \pgfkeyssetvalue{/pgfplots/table/@cell content}{\rotatebox{90}{##1}\\}%
            \else
              \ifnum\pgfplotstablecol=1
                \pgfkeyssetvalue{/pgfplots/table/@cell content}{\hspace*{8mm}##1&}%
              \else
                \pgfkeyssetvalue{/pgfplots/table/@cell content}{\rotatebox[origin=lb]{90}{##1}&}%
              \fi
            \fi
          }
        },
        every last row/.style={after row=\bottomrule},
      ]\correctnessstats
    \end{center}
    \begin{center}
      \begin{tabular}{llll}
        \CIRCLE{} Detected & \LEFTcircle{} Different Error &
        \Circle{} Not Detected & \textsuperscript {\textdagger}SafeFFI-specific Error \\
        \multicolumn{4}{c}{INTERCEPT = Interception-based Detection}
      \end{tabular}
    \end{center}
    \vspace*{-5mm}
\end{table}
\CommentCameraReady{For vulnerablity-table: Add information about the type of vulnerability; Differentiate SafeFFI-specific Errors; Question: can SafeFFI detect those vulnerabilities with out re-inserted cg-based checks?}
\CommentCameraReady{JH: vulnerability-table: the intercept column does not really get a discussion in text, does it? and why does it use a different marker style that is not in the legend?}

\begin{figure*}
  \centering
  \captionsetup{skip=2pt}
  \input{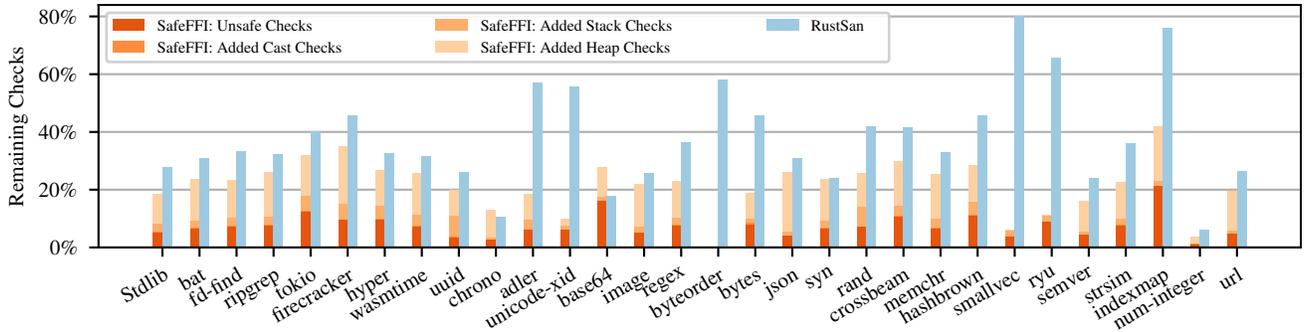}
  \caption{Sanitizer checks after elision by \ourApproach and RustSan, respectively, as fractions of original checks for individual packages. For \ourApproach, checks are subdivided into remaining unsafe checks and added cast, stack and heap checks.}
  \label{fig:elisions}
\end{figure*}

\subsection{Effectiveness}
\label{sec:eval:effectivness}
In this section, we provide empirical evidence that \ourApproach is effective in reducing the number of \hwasan checks (\ref{rq:elided_checks}) and, consequently, the run-time overhead of the sanitizer (\ref{rq:runtime}).\
We further compare \ourApproach against RustSan. As mentioned in \autoref{sec:correctness}, ERASan incorrectly removes necessary checks, leading to a skewed number of elided checks. Thus, we exclude ERASan from further comparisons.
We use the same crates (Rust's term for packages or libraries) as RustSan for the evaluation.
Note that for RustSan, neither the evaluation scripts nor the tested versions of the crates are available, thus we chose a common version that both \ourApproach and RustSan can compile with their respective tool chains.
The crates Rocket and RustPython are missing because they contain a compilation error in the version compatible with Rust 1.52.0 and we bundled the libstd tests into a single crate.
\CommentCameraReady{JH: maybe one sentence summarizing which (type of) applications/crates are being tested? how much unsafe code/C/C++ do they include?}
Elided checks are measured at LLVM IR level. To prevent the standard library from overshadowing the results of smaller crates, we only count checks outside the standard library and list the standard library results separately.

\paragraph{Elided Checks.}

\autoref{fig:elisions} shows \ourApproach's capability to elide sanitizer checks. 
\ourApproach retains on average $\SafeFFIUnsafeChecksAvg \%$ of the \gls{HWASAN} checks because they vet an unsafe pointer.\tdprioB{turning those numbers around sounds better: we elide 90\% of the checks} Then, cast checks (cf. \autoref{sec:concept_pointer-casts-boundary}) and stack checks (cf. \autoref{sec:interprocedural-invariant}) are added to ensure spatial and stack-temporal memory safety for safe pointers. After these checks, on average $\SafeFFIUnsafeAndAddedAvg \%$ of the checks remain. Adding the optional heap checks for \FreeDuringScope vulnerabilities~(cf. \autoref{sec:heap-temporal}) increases the average number of remaining checks to $\SafeFFIFullChecksAvg \%$. 
\CommentCameraReady{All additionally added checks grow proportionally with the total number of checks, showing that within the dataset, there are no patterns that significantly change the elision rate.really? double check by printing percentages of added checks in appendix tables}

Compared to RustSan (blue bars in \autoref{fig:elisions}), \ourApproach{} with full temporal safety elides on weighted average $\SafeFFIVsRustSanAvg \%$ \emph{more} checks, providing a significant improvement over the current state of the art. On closer examination, \ourApproach performs better on \ElisionNumSafeFFIBetter{} benchmarks and worse on \ElisionNumRustSanBetter{} benchmarks, exceeding $5 \%$ difference on \ElisionNumSafeFFIBetterFivePercent{} of \ElisionNumSafeFFIBetter{} benchmarks and \ElisionNumRustSanBetterFivePercent{} of \ElisionNumRustSanBetter{} benchmarks respectively.\CommentCameraReady{Nummern hier in dem Text 3x checken nachdem die Tabellen final sind}
\tdprioB{TL: smallvec fits the same argument, should we include it again?}
We observe especially large difference ($>20 \%$) on bytes, ryu, indexmap (in favor of \ourApproach) and base64 (in favor of RustSan). bytes and indexmap are data structures with unsafe pointers that alias with safe views returned by the API. Similarly, in ryu crucial internal buffers make use of unsafe. In these cases, \ourApproach already performs better in isolation and the advantage should even increase in dependents of these libraries. Furthermore, as ryu contains no heap allocations, \ourApproach does not need to add any heap checks, which further increases the savings.
In contrast, on base64 \ourApproach performs significantly worse.
The reason is that base64 is lowered into many dynamic GEP statements, which we conservatively overapproximate (cf.~\autoref{sec:architecture:pointer-type-analysis}).  
Note that \ourApproach and RustSan cannot operate in the exact same environment---however, it is unlikely that the differences affect elision rates. 

\begin{figure*}
  \centering
  \captionsetup{skip=2pt}
  \input{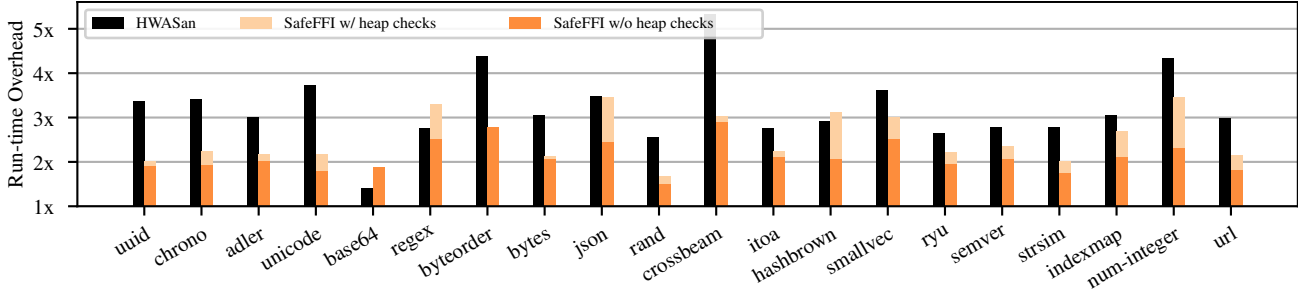}
  \caption{Run-time overhead of vanilla HWAsan, \ourApproach (including cast and stack checks), and \ourApproach with added heap checks compared to baseline run-time without sanitizer.}
  \label{fig:runtime_overhead}
\end{figure*}

\paragraph{Run-time Performance.}
We have shown that \ourApproach is superior in eliding checks. However, not all checks are equally important for the run-time, checks on hot paths through the program disproportionately impact the run-time~\cite{HighSystemCodeSecurity2015}. Therefore, we also evaluate the run-time on several benchmarks. \autoref{fig:runtime_overhead} shows the run-time overhead of \gls{HWASAN} and \ourApproach on the benchmarks. \gls{HWASAN} imposes an average runtime overhead of $\HwasanRtOhAvg \times$ (median: $\HwasanRtOhMedian \times$) compared to an uninstrumented binary. \ourApproach without heap-temporal-safety reduces this overhead to $\SafeFFINoneRtOhAvg \times$ (median: $\SafeFFINoneRtOhMedian \times$) and \ourApproach with full temporal-safety to $\SafeFFIRtOhAvg \times$ (median: $\SafeFFIRtOhMedian \times$). 
Considering each benchmark individually, \ourApproach improves the run-time by at least $10\%$ across the board, except for four crates. 
Base64 contains a lot of dynamic GEP instructions, which \ourApproach conservatively overapproximates, as explained above.
For json and hashbrown, the additional heap checks required to detect \FreeDuringScope vulnerabilities (see \autoref{sec:heap-temporal}) limit the achievable performance improvement, without added heap checks \ourApproach significantly reduces the run-time overhead.
For regex, none of the added checks are proportionally bigger compared to other crates, therefore, we assume that some inserted checks are on hot paths in the benchmark execution.

Only base64 $\SafeFFINoneNoImprov/\NumRtBenchmarks$ for \ourApproach None and $\SafeFFINoImprov/\NumRtBenchmarks$ for \ourApproach not exceeding a 10\% improvement in overhead.\tdob{Explain more weird effects: bad performance for base and regex (also json, hashbrown, smallvec)}
\tdjh{i think this is hard to parse, I would write "only 1 of 20 apps". and then, still, this looks a bit like trying hard to make it sound good, picking stuff very selectively. i get it but I think reviewers see past this}
Unfortunately, we could not include benchmarking data for RustSan because the cargo benchmark harness compiled with RustSan consistently raises segmentation faults during startup. %
As target versions and benchmarking scripts are not included in RustSan's published code, we also cannot perform a like-for-like comparioson against the numbers reported in their paper.  
Deducing from our comparison of elided checks with their reported results, we suspect that they benchmarked without compiling an instrumented version of the Rust standard library which could introduce false negatives and thus distorts the comparison.\CommentCameraReady{Most of the benchmarks were measured with cargo bench. Does not filter outliers (with test pc not being completely isolated). Might not be the best benchmarker to use. Criterion might be better}

\pgfplotstableread[col sep=comma]{data/runtime_stats.csv}\runtimestats
\pgfplotstableset{
    runtime_style/.style={
    columns/Competitor Benchmarks/.style={
        column name={Benchmark},
        string type,
        column type={l},
    },
    columns/baseline_runtime(ns)/.style={
        column name={Baseline (ns)},
        column type={r},
        precision=2,
        fixed zerofill,
    },
    columns/hwasan_overhead/.style={
        column name={HWASan},
        column type={r},
        precision=2,
        fixed zerofill,
        postproc cell content/.append style={
            /pgfplots/table/@cell content/.add={}{$\times$}
        },
    },
    columns/safeffi-none_overhead/.style={
        column name={SafeFFI w/o heap checks},
        column type={r},
        precision=2,
        fixed zerofill,
        postproc cell content/.append style={
            /pgfplots/table/@cell content/.add={}{$\times$}
        },
    },
    columns/safeffi-tim-callgraph_overhead/.style={
        column name={SafeFFI w/ heap checks},
        column type={r},
        precision=2,
        fixed zerofill,
        postproc cell content/.append style={
            /pgfplots/table/@cell content/.add={}{$\times$}
        },
    },
    columns/RustSan Overhead/.style={
        column name={RustSan},
        column type={||r},
        precision=2,
        fixed zerofill,
        postproc cell content/.append style={
            /pgfplots/table/@cell content/.add={}{$\times$}
        },
    },
    columns/ERASan-rawptr Overhead/.style={
        column name={ERASAN-rawptr (Reported)},
        column type={r},
        precision=2,
        fixed zerofill,
        postproc cell content/.append style={
            /pgfplots/table/@cell content/.add={}{$\times$}
        },
    },
    every head row/.style={
        before row=\toprule,
        after row=\midrule
    },
    every last row/.style={after row=\bottomrule},
    }
}

\subsection{Robustness in \gls{MLA} Scenarios}
\label{sec:eval:mla}
Our experiments on correctness (\autoref{sec:correctness}) and effectiveness (\autoref{sec:eval:effectivness}) contain real-world \glspl{MLA} in the benchmark sets, e.g., bat depending on libgit2, rusqlite~(CVE-2021-45713) on libsqlite3, or xcb~(RUSTSEC-2020-0097) on libxcb.
Since we did not encounter \gls{FFI}-related compilation issues nor false positives/negatives during execution, this is evidence that \ourApproach works as designed in \gls{MLA} scenarios.
\tdprioC{JH: no false negatives sounds a bit too general IMO}
\CommentCameraReady{print crate names in known-vulnerabilities-table}
To be confident in the robustness of \ourApproach, we created a systematic test set of minimal test cases that cover all common FFI interactions between C and Rust that we could conceive of. It covers the following dimensions:
\begin{enumerate}
  \item Allocation: Global, C stack/heap, Rust stack/heap;
  \item Deallocation: Global, C stack/heap, Rust stack/heap;
  \item Pointer invalidation: pointer arithmetic, deallocation, invalid pointer crafting, or no invalidation (benign test);
  \item Control flow permutation: interleaving of pointer casts, invalidations, and dereferences;
\end{enumerate}
We evaluated \ourApproach on all meaningful combinations of these dimensions leading to 45 distinct test cases (35 with vulnerabilities, 10 without).
All tests are included in the artifact.
\ourApproach is robust against all of those cases, meaning that it does not raise any false positives or false negatives.

\subsection{Compile-Time Performance}
\label{sec:compile-time-performance}

\begin{figure*}[t]
  \centering
  \captionsetup{skip=2pt}
  \input{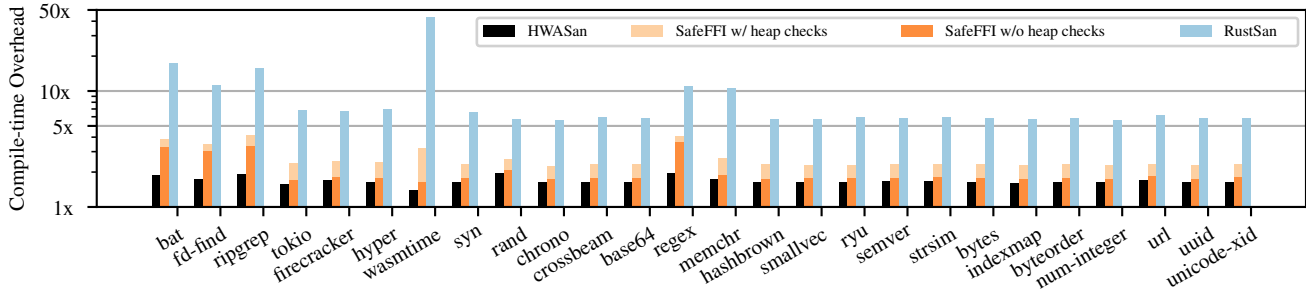}
  \caption{Compilation-time overhead of vanilla HWAsan, \ourApproach (including cast and stack checks), and \ourApproach with added heap checks compared to baseline compilation-time without sanitizer. Note that the scale is logarithmic to accommodate large overheads of RustSan.}
  \label{fig:compilation}
\end{figure*}

\autoref{fig:compilation} shows the compile-time overhead of \gls{HWASAN}, \ourApproach and RustSan, answering \ref{rq:compiletime}. Note that we chose the respective toolchain versions as a baseline for a fair comparison. On average, \gls{HWASAN} induces a compile-time overhead of $\HwasanCtOhAvg \times$ compared to the baseline. \ourApproach without heap-temporal-safety increases the overhead to $\SafeFFINoneCtOhAvg \times$. \ourApproach with full temporal-safety further increases the compile-time overhead to $\SafeFFICtOhAvg \times$.
The compile-time memory overhead of \ourApproach is reasonable, with an average of $\SafeFFICmOhAvg \times$ compared to the baseline and within measurement deviation compared to \gls{HWASAN}.
RustSan adds a compile-time overhead of $\RustSanCtOhAvg \times$ on average, with outliers reaching an overhead of up to $\RustSanCtOhMax \times$, because it performs a whole-program points-to analysis to identify safe pointers that alias raw pointers. For the same reason, RustSan also induces a significant memory overhead during compilation of $\RustSanCmOhAvg \times$ on average because bigger crates like wasmtime consuming up to $\RustSanCmOhMax \times$ more RAM.

\pgfplotstableread[col sep=comma]{data/compilation_stats.csv}\compilestats
\pgfplotstableset{
    compilation_style/.style={
      every col no 3/.append style={postproc cell content/.append style={/pgfplots/table/@cell content/.add={}{$\times$}}},
      every col no 4/.append style={postproc cell content/.append style={/pgfplots/table/@cell content/.add={}{$\times$}}},
      every col no 5/.append style={postproc cell content/.append style={/pgfplots/table/@cell content/.add={}{$\times$}}},
      every col no 6/.append style={postproc cell content/.append style={/pgfplots/table/@cell content/.add={}{$\times$}}},
      every col no 7/.append style={postproc cell content/.append style={/pgfplots/table/@cell content/.add={}{$\times$}}},
      every col no 8/.append style={postproc cell content/.append style={/pgfplots/table/@cell content/.add={}{$\times$}}},
      every col no 9/.append style={postproc cell content/.append style={/pgfplots/table/@cell content/.add={}{$\times$}}},
      every col no 10/.append style={postproc cell content/.append style={/pgfplots/table/@cell content/.add={}{$\times$}}},
      columns/{benchmark}/.append style={
        column name={Benchmark},
        string type,
        column type={l}, %
      },
      columns/{baseline_compile_time_s}/.style={
        column name={Time (s)},
        column type={r},
        precision=2,
        fixed zerofill,
      },
      columns/{baseline_memory_mb}/.style={
        column name={Memory (MB)},
        column type={r},
        precision=2,
        fixed zerofill,
      },
      columns/{hwasan_compile_time_overhead}/.style={
        column name={Time OH},
        column type={r},
        precision=2,
        fixed zerofill,
      },
      columns/{hwasan_memory_overhead}/.style={
        column name={Mem OH},
        column type={r},
        precision=2,
        fixed zerofill,
      },
      columns/{safeffi-none_compile_time_overhead}/.style={
        column name={Time OH},
        column type={r},
        precision=2,
        fixed zerofill,
      },
      columns/{safeffi-none_memory_overhead}/.style={
        column name={Mem OH},
        column type={r},
        precision=2,
        fixed zerofill,
      },
      columns/{safeffi-tim-callgraph_compile_time_overhead}/.style={
        column name={Time OH},
        column type={r},
        precision=2,
        fixed zerofill,
      },
      columns/{safeffi-tim-callgraph_memory_overhead}/.style={
        column name={Mem OH},
        column type={r},
        precision=2,
        fixed zerofill,
      },
      columns/{rustsan_compile_time_overhead}/.style={
        column name={Time OH},
        column type={r},
        precision=2,
        fixed zerofill,
      },
      columns/{rustsan_memory_overhead}/.style={
        column name={Mem OH},
        column type={r},
        precision=2,
        fixed zerofill,
      },
      every head row/.style={
          before row={
              \toprule
                    \multicolumn{1}{c}{} & \multicolumn{2}{c}{\textbf{Baseline}} & \multicolumn{2}{c}{\textbf{HWASan}} & \multicolumn{2}{c}{\textbf{SafeFFI w/o heap checks}} & \multicolumn{2}{c}{\textbf{SafeFFI w/ heap checks}} & \multicolumn{2}{c}{\textbf{RustSan}}\\
                    \cmidrule(lr){2-3} \cmidrule(lr){4-5} \cmidrule(lr){6-7} \cmidrule(lr){8-9} \cmidrule(lr){10-11}
          }, 
          after row=\midrule
      },
      every last row/.style={after row=\bottomrule},
    }
}

\section{Related Work}
\label{sec:related-work}

\paragraph{Run-time Check Reduction for Rust.}
Rust for Morello~\cite{RustMorelloAlwaysOn2023} executes Rust programs on the CHERI-Architecture\cite{CHERIHybridCapabilitySystem2015} for hardware-enforced memory safety (not commercially available). They elide software bounds checks emitted by the Rust compiler, but find little benefit due to compiler optimizations.

All three approaches annotate pointers during MIR lowering so that safe and \raw pointers remain distinguishable at the LLVM IR level, with the goal to elide checks for \safe pointers.
The key difference is that RustSan and ERASan rely on whole-program static value-flow analysis to compute points-to sets, while \ourApproach annotates function parameters and return values with pointer types, allowing for efficient local reasoning.
This design significantly reduces compile-time overhead and scales better to large code bases.

Moreover, this difference makes \ourApproach significantly more effective at eliding checks.
This happens not only because the static analysis tends to overapproximate, but also for a fundamental reason: Rust code that interacts with \unsafe or foreign code often uses patterns where \safe pointers are derived from \raw pointers \emph{or vice versa}.
RustSan and ERASan need to insert checks for every pointer which shares the same points-to set as the \raw pointer, inhibiting the elision of checks for \safe pointers in this case.
Our approach, on the other hand, benefits from the chance that the actual number of \raw pointer interactions might be small or just unidirectional, because \ourApproach just inserts a check at casts from \raw pointers to \safe pointers.
This makes \ourApproach especially suitable for scenarios involving \unsafe code, e.g., fuzzing Rust crates that implement efficient datastructures or protecting \glspl{MLA}.
While not explicitly stated, we deem RustSan's and ERASan's concept also to be applicable to \glspl{MLA}.

Our concept of hoisting checks to locations of casts and other \safe pointer creations also leads to superior debuggability because \ourApproach fails earlier and points developers directly to a violation of the Rust's requirements for \safe pointers at the boundary to \unsafe or foreign code.
Theoretically, our check hoisting concept introduces a risk of missing \FreeDuringScope temporal vulnerabilities (see \autoref{sec:heap-temporal}), an issue for which we provide a solution for single-threaded programs by optionally inserting additional sanitizer checks at potential heap deallocation operations.
\ourApproach still outperforms RustSan and ERASan on the evaluated metrics in most cases.
Those two approaches conceptually avoid this risk by retaining all sanitizer checks at pointer dereference locations if the pointer might alias with a \raw pointer.

However, interestingly, RustSan and ERASan show worse ability to retain the sanitizer's detection capabilities on our tests with real-world vulnerabilities (\autoref{tab:knownvulns}).
ERASan in particular suffers from implementation bugs that cause false negatives. 
Both implement a blacklist approach for identifying \raw pointers, potentially leading to false negatives if the type annotation is incomplete.
\ourApproach uses a whitelist approach instead, validating that \emph{every} pointer has a type and overapproximating \raw pointers where the implementation is incomplete---which should be accounted for because the Rust compiler's internals for lowering MIR to LLVM IR are very complex.
Overall, our implementation is more tightly integrated with the Rust compilation toolchain without the need to link and analyze the program externally.
This should increase the chances of adoption.

\paragraph{General Sanitizer Optimizations.}
Prior work on C/C++ includes ASAP~\cite{HighSystemCodeSecurity2015} which removes checks on hot paths, trading security for performance. Moll et al.~\cite{BoundsCheckHoisting2014}, hoist bounds checks in loops. ASan\texttt{-}\texttt{-}~\cite{DebloatingAddressSanitizer2022} uses lightweight static analysis and SanRazor~\cite{SANRAZORReducingRedundant2021} a combination of static and dynamic analysis to reduce redundant checks. These are complementary to \ourApproach and could be applied to raw pointers in Rust or \Ccpp parts of \glspl{MLA}.

\paragraph{Runtime Isolation for Rust.}
Several proposed approaches isolate safe Rust from unsafe code, e.g., Sandcrust~\cite{SandcrustAutomaticSandboxing2017}, Fidelius Charm~\cite{FideliusCharmIsolating2018}, XRust~\cite{SecuringUnsafeRust2020}, relying on developer annotations or hardware/OS support for process isolation. 
TRust~\cite{TRUSTCompilationFrameworka}, PRKUSafe~\cite{PKRUsafeAutomaticallyLocking2022} and Gülmez et al.\cite{FriendFoeExploring2023} instead use Intel's \gls{MPK} to separate safe and unsafe objects in different memory regions.
Galeed~\cite{KeepingSafeRust2021} and Omniglot~\cite{Omniglot2025} explicitly address cross-language attacks\cite{CrossLanguageAttacks2022} by isolating Rust from C code, relying on hardware features for in-process isolation.
As opposed to those isolation approaches that only inhibit the spread of memory vulnerabilities from unsafe code to safe Rust code, \ourApproach prevents memory vulnerabilities in the first place and therefore enables better debugging, without requiring source or OS changes.

\paragraph{Static Analyzers for Memory Safety in Rust.}
Tools like MIRChecker~\cite{MirCheckerDetectingBugs}, Rudra~\cite{RudraFindingMemory2021}, Yuga~\cite{Yuga2024}, and SafeDrop~\cite{SafeDropDetectingMemory2023} use static analysis to detect memory bugs in Rust, while FFIChecker~\cite{DetectingCrosslanguageMemory2022} and CRust~\cite{CRustCrossLang2022} focus on FFI safety. These approaches suffer from false positives and excessive compile-time overhead, whereas \ourApproach is a dynamic analysis based on sanitizers resulting in high precision at the cost of run-time overheads which \ourApproach significantly reduces for Rust code.

\section{Conclusion}
We presented \ourApproach, a novel approach to hoist sanitizer checks in Rust programs and \glspl{MLA} to reduce the run-time overhead of sanitizers.
Compared to existing approaches, \ourApproach only uses funcion-local reasoning to hoist checks instead of depending on whole-program static points-to analysis which can be expensive and error-prone.
Our approach for hoisting checks to locations of casts from \raw pointers to \safe pointers shows improved performance compared to existing methods, especially for Rust programs that frequently interact with unsafe code or FFI code.
We showed practicality and effectiveness in reducing the total number of sanitizer checks for popular Rust libraries by $58\%-98\%$, resulting in a reduction of the run-time overhead of \gls{HWASAN} from $\HwasanRtOhAvg\times$ to $\SafeFFIRtOhAvg\times$ on average.\question{is that a good way to summarize the numbers?}\CommentCameraReady{double check numbers (58=indexmap, 98=num-integer)}
Our experiments with existing real-world vulnerabilities show that \ourApproach not only maintains the sanitizer's detection capabilities but also improves debuggability by failing closer to the root cause of memory safety violations in the interaction between safe rust and unsafe Rust or foreign code.
We showed that \ourApproach can be implemented in a very modular way such that in the future it can be adapted for different memory sanitizers that choose different tradeoffs like SoftboundCETS~\cite{NZMZ09,NZMZ10} for more security.

\cleardoublepage
\section*{Ethical Considerations}
\ourApproach is a purely defensive tool to improve the efficiency of memory sanitizers in Rust programs. Enabling more software to run with memory sanitization should have a net benefit in reducing vulnerabilities. A potential risk is that benign executions could be terminated although the observed memory error would not have had negative effects, but we believe that this risk is outweighed by the benefits.

All vulnerabilities used for evaluation in this work (see Table \ref{tab:knownvulns}) were already publicly known, and we did not try to discover new vulnerabilities.

\section*{Open Science}
We comply with the open science policy by releasing the \ourApproach prototype as open source and for artifact evaluation.
This includes our modifications to rustc and LLVM, test sets, benchmarks and the corresponding scripts to build and execute them with \ourApproach.
An anonymized version for review can be found on Zenodo:\\\url{https://tinyurl.com/safeffi-usenix}

The final version will be published on Github upon acceptance.\CommentCameraReady{TODO: prepare github release}

\bibliographystyle{plain}
\bibliography{SafeFFI,SafeFFI.bib_additionsJH,SafeFFI.bib_additionsTL}

\begin{thebibliography}{10}

\bibitem{FideliusCharmIsolating2018}
Hussain~MJ Almohri and David Evans.
\newblock Fidelius {{Charm}}: {{Isolating Unsafe Rust Code}}.
\newblock In {\em Proceedings of the {{Eighth ACM Conference}} on {{Data}} and
  {{Application Security}} and {{Privacy}}}, pages 248--255. ACM.

\bibitem{RudraFindingMemory2021}
Yechan Bae, Youngsuk Kim, Ammar Askar, Jungwon Lim, and Taesoo Kim.
\newblock Rudra: {{Finding Memory Safety Bugs}} in {{Rust}} at the {{Ecosystem
  Scale}}.
\newblock In {\em Proceedings of the {{ACM SIGOPS}} 28th {{Symposium}} on
  {{Operating Systems Principles}}}, {{SOSP}} '21, pages 84--99. Association
  for Computing Machinery.

\bibitem{TRUSTCompilationFrameworka}
Inyoung Bang, Martin Kayondo, HyunGon Moon, and Yunheung Paek.
\newblock $\{$TRust$\}$: A compilation framework for in-process isolation to
  protect safe rust against untrusted code.
\newblock In {\em 32nd USENIX Security Symposium (USENIX Security 23)}, pages
  6947--6964, 2023.

\bibitem{chen2023mtsan}
Xingman Chen, Yinghao Shi, Zheyu Jiang, Yuan Li, Ruoyu Wang, Haixin Duan, Haoyu
  Wang, and Chao Zhang.
\newblock {MTSan}: A feasible and practical memory sanitizer for fuzzing {COTS}
  binaries.
\newblock In {\em USENIX Security}. USENIX Association, 2023.

\bibitem{RustSanRetrofittingAddressSanitizer2024}
Kyuwon Cho, Jongyoon Kim, Kha~Dinh Duy, Hajeong Lim, and Hojoon Lee.
\newblock $\{$RustSan$\}$: Retrofitting $\{$AddressSanitizer$\}$ for efficient
  sanitization of rust.
\newblock In {\em 33rd USENIX Security Symposium (USENIX Security 24)}, pages
  3729--3746, 2024.

\bibitem{SafeDropDetectingMemory2023}
Mohan Cui, Chengjun Chen, Hui Xu, and Yangfan Zhou.
\newblock Safedrop: Detecting memory deallocation bugs of rust programs via
  static data-flow analysis.
\newblock {\em ACM Transactions on Software Engineering and Methodology},
  32(4):1--21, 2023.

\bibitem{RustReferenceSafety}
The Rust~Project Developers.
\newblock {Primitive Type Reference - Safety}.
\newblock Section of "The Rust Standard Library" documentation.
  \url{{https://doc.rust-lang.org/std/primitive.reference.html#safety}}.

\bibitem{RustReferenceTypeCoercions}
The Rust~Project Developers.
\newblock {Type Coercions}.
\newblock Section of "The Rust Reference".
  \url{https://doc.rust-lang.org/reference/type-coercions.html}.

\bibitem{FriendFoeExploring2023}
Merve Gülmez, Thomas Nyman, Christoph Baumann, and Jan~Tobias Mühlberg.
\newblock Friend or foe inside? exploring in-process isolation to maintain
  memory safety for unsafe rust, 2023.
\newblock \url{https://arxiv.org/abs/2306.08127}.

\bibitem{RustMorelloAlwaysOn2023}
Sarah Harris, Simon Cooksey, Michael Vollmer, and Mark Batty.
\newblock Rust for {{Morello}}: {{Always-On Memory Safety}}, {{Even}} in
  {{Unsafe Code}} ({{Experience Paper}}).
\newblock In {\em
  {{DROPS-IDN}}/v2/Document/10.4230/{{LIPIcs}}.{{ECOOP}}.2023.39}.
  Schloss-Dagstuhl - Leibniz Zentrum für Informatik.

\bibitem{projectzero-zerodays}
Ben Hawkes.
\newblock 0day ``in the wild'', 2019.
\newblock \url{https://googleprojectzero.blogspot.com/p/0day.html}.

\bibitem{HZH23}
Konrad Hohentanner, Philipp Zieris, and Julian Horsch.
\newblock {CryptSan}: Leveraging {ARM} {Pointer} {Authentication} for memory
  safety in {C}/{C++}.
\newblock In {\em SAC}. ACM, 2023.

\bibitem{CRustCrossLang2022}
Shuang Hu, Baojian Hua, Lei Xia, and Yang Wang.
\newblock {CRUST:} towards a unified cross-language program analysis framework
  for rust.
\newblock In {\em 22nd {IEEE} International Conference on Software Quality,
  Reliability and Security, {QRS} 2022, Guangzhou, China, December 5-9, 2022},
  pages 970--981. {IEEE}, 2022.

\bibitem{Intel-PC}
{Intel Corp.}
\newblock Pointer checker, 2021.
\newblock
  \url{https://www.intel.com/content/www/us/en/docs/cpp-compiler/developer-guide-reference/2021-10/pointer-checker.html}.

\bibitem{PKRUsafeAutomaticallyLocking2022}
Paul Kirth, Mitchel Dickerson, Stephen Crane, Per Larsen, Adrian Dabrowski,
  David Gens, Yeoul Na, Stijn Volckaert, and Michael Franz.
\newblock {{PKRU-safe}}: Automatically locking down the heap between safe and
  unsafe languages.
\newblock In {\em Proceedings of the {{Seventeenth European Conference}} on
  {{Computer Systems}}}, pages 132--148. ACM.

\bibitem{RustBookOwnershipChapter}
Steve Klabnik, Carol Nichols, and Chris Krycho.
\newblock {Understanding Ownership}.
\newblock Chapter 4 of "The Rust Programming Language".
  \url{https://doc.rust-lang.org/book/ch04-00-understanding-ownership.html}.

\bibitem{kroes2018delta}
Taddeus Kroes, Koen Koning, Erik van~der Kouwe, Herbert Bos, and Cristiano
  Giuffrida.
\newblock {Delta pointers}: Buffer overflow checks without the checks.
\newblock In {\em EuroSys}. ACM, 2018.

\bibitem{SandcrustAutomaticSandboxing2017}
Benjamin Lamowski, Carsten Weinhold, Adam Lackorzynski, and Hermann Härtig.
\newblock Sandcrust: {{Automatic}} sandboxing of unsafe components in rust.
\newblock In {\em Proceedings of the 9th {{Workshop}} on {{Programming
  Languages}} and {{Operating Systems}}}, pages 51--57. ACM.

\bibitem{PACMem}
Yuan Li, Wende Tan, Zhizheng Lv, Songtao Yang, Mathias Payer, Ying Liu, and
  Chao Zhang.
\newblock {PACMem}: Enforcing spatial and temporal memory safety via {ARM}
  {P}ointer {A}uthentication.
\newblock In {\em CCS}. ACM, 2022.

\bibitem{DetectingCrosslanguageMemory2022}
Zhuohua Li, Jincheng Wang, Mingshen Sun, and John C.~S. Lui.
\newblock Detecting {{Cross-language Memory Management Issues}} in~{{Rust}}.
\newblock In Vijayalakshmi Atluri, Roberto Di~Pietro, Christian~D. Jensen, and
  Weizhi Meng, editors, {\em Computer {{Security}} – {{ESORICS}} 2022},
  Lecture {{Notes}} in {{Computer Science}}, pages 680--700. Springer Nature
  Switzerland.

\bibitem{MirCheckerDetectingBugs}
Zhuohua Li, Jincheng Wang, Mingshen Sun, and John~CS Lui.
\newblock Mirchecker: detecting bugs in rust programs via static analysis.
\newblock In {\em Proceedings of the 2021 ACM SIGSAC conference on computer and
  communications security}, pages 2183--2196, 2021.

\bibitem{CAMP}
Zhenpeng Lin, Zheng Yu, Ziyi Guo, Simone Campanoni, Peter Dinda, and Xinyu
  Xing.
\newblock {CAMP}: Compiler and allocator-based heap memory protection.
\newblock In {\em USENIX Security}. USENIX Association, 2024.

\bibitem{SecuringUnsafeRust2020}
Peiming Liu, Gang Zhao, and Jeff Huang.
\newblock Securing unsafe rust programs with {{XRust}}.
\newblock In {\em Proceedings of the {{ACM}}/{{IEEE}} 42nd {{International
  Conference}} on {{Software Engineering}}}, pages 234--245. ACM.

\bibitem{CrossLanguageAttacks2022}
Samuel Mergendahl, Nathan Burow, and Hamed Okhravi.
\newblock Cross-{{Language Attacks}}.
\newblock In {\em Proceedings 2022 {{Network}} and {{Distributed System
  Security Symposium}}}. Internet Society.

\bibitem{miller2019trends}
Matt Miller.
\newblock Trends, challenges, and strategic shifts in the software
  vulnerability mitigation landscape.
\newblock In {\em BlueHat IL}. Microsoft Security Response Center, 2019.

\bibitem{ERASanEfficientRust2024}
Jiun Min, Dongyeon Yu, Seongyun Jeong, Dokyung Song, and Yuseok Jeon.
\newblock {{ERASan}}: {{Efficient Rust Address Sanitizer}}.
\newblock In {\em 2024 {{IEEE Symposium}} on {{Security}} and {{Privacy}}
  ({{SP}})}, pages 4053--4068.

\bibitem{CWETop10KEV}
{MITRE Corp.}
\newblock 2024 {CWE} top 10 kev weaknesses, 2024.
\newblock \url{https://cwe.mitre.org/top25/archive/2024/2024_kev_list.html}.

\bibitem{CWETop25}
{MITRE Corp.}
\newblock 2024 {CWE} top 25 most dangerous software weaknesses, 2024.
\newblock \url{https://cwe.mitre.org/top25/archive/2024/2024_cwe_top25.html}.

\bibitem{BoundsCheckHoisting2014}
Simon Moll, Henrique Nazaré, Gustavo~Vieira Machado, and Raphael~Ernani
  Rodrigues.
\newblock Bounds {{Check Hoisting}} for {{AddressSanitizer}}.
\newblock In Fernando~Magno Quintão~Pereira, editor, {\em Programming
  {{Languages}}}, pages 47--61. Springer International Publishing.

\bibitem{NZMZ09}
Santosh Nagarakatte, Jianzhou Zhao, Milo M.~K. Martin, and Steve Zdancewic.
\newblock Softbound: Highly compatible and complete spatial memory safety for
  {C}.
\newblock In {\em {PLDI}}. {ACM}, 2009.

\bibitem{NZMZ10}
Santosh Nagarakatte, Jianzhou Zhao, Milo~M.K. Martin, and Steve Zdancewic.
\newblock {CETS}: Compiler enforced temporal safety for {C}.
\newblock In {\em {ISMM}}. {ACM}, 2010.

\bibitem{Yuga2024}
Vikram Nitin, Anne Mulhern, Sanjay Arora, and Baishakhi Ray.
\newblock Yuga: Automatically detecting lifetime annotation bugs in the rust
  language.
\newblock {\em IEEE Trans. Softw. Eng.}, 50(10):2602–2613, October 2024.

\bibitem{OBZH24}
Benjamin Orthen, Oliver Braunsdorf, Philipp Zieris, and Julian Horsch.
\newblock {SoftBound+CETS} revisited: More than a decade later.
\newblock In {\em EuroSec}. ACM, 2024.

\bibitem{ExploitingMixedBinaries2021}
Michalis Papaevripides and Elias Athanasopoulos.
\newblock Exploiting mixed binaries.
\newblock {\em ACM Transactions on Privacy and Security (TOPS)}, 24(2):1--29,
  2021.

\bibitem{KeepingSafeRust2021}
Elijah Rivera, Samuel Mergendahl, Howard Shrobe, Hamed Okhravi, and Nathan
  Burow.
\newblock Keeping {{Safe Rust Safe}} with {{Galeed}}.
\newblock In {\em Annual {{Computer Security Applications Conference}}},
  {{ACSAC}}, pages 824--836. Association for Computing Machinery.

\bibitem{Omniglot2025}
Leon Schuermann, Jack Toubes, Tyler Potyondy, Pat Pannuto, Mae Milano, and Amit
  Levy.
\newblock Building bridges: Safe interactions with foreign languages through
  omniglot.
\newblock In Lidong Zhou and Yuanyuan Zhou, editors, {\em 19th {USENIX}
  Symposium on Operating Systems Design and Implementation, {OSDI} 2025,
  Boston, MA, USA, July 7-9, 2025}, pages 595--613. {USENIX} Association, 2025.

\bibitem{asan}
Konstantin Serebryany, Derek Bruening, Alexander Potapenko, and Dmitry Vyukov.
\newblock {AddressSanitizer}: A fast address sanity checker.
\newblock In {\em {USENIX} {ATC}}. {USENIX} Association, 2012.

\bibitem{hwasan}
Kostya Serebryany, Evgenii Stepanov, Aleksey Shlyapnikov, Vlad Tsyrklevich, and
  Dmitry Vyukov.
\newblock Memory tagging and how it improves {C/C++} memory safety.
\newblock arxiv:1802.09517, {Google LLC}, 2018.

\bibitem{c2rustbench}
Melih Sirlanci, Carter Yagemann, and Zhiqiang Lin.
\newblock {C2RUST-BENCH:} {A} minimized, representative dataset for c-to-rust
  transpilation evaluation.
\newblock {\em CoRR}, abs/2504.15144, 2025.

\bibitem{SLR+19}
Dokyung Song, Julian Lettner, Prabhu Rajasekaran, Yeoul Na, Stijn Volckaert,
  Per Larsen, and Michael Franz.
\newblock {SoK}: Sanitizing for security.
\newblock In {\em {S\&P}}. {IEEE}, 2019.

\bibitem{chromium-memsafety}
{The Chromium Developers}.
\newblock Memory safety.
\newblock \url{https://www.chromium.org/Home/chromium-security/memory-safety/}.

\bibitem{vintila-mset2025-sos}
Emanuel Vintila, Philipp Zieris, and Julian Horsch.
\newblock {Evaluating the Effectiveness of Memory Safety Sanitizers}.
\newblock In {\em {2025 IEEE Symposium on Security and Privacy (SP)}}, pages
  88--88, Los Alamitos, CA, USA, May 2025. IEEE Computer Society.

\bibitem{HighSystemCodeSecurity2015}
Jonas Wagner, Volodymyr Kuznetsov, George Candea, and Johannes Kinder.
\newblock High {{System-Code Security}} with {{Low Overhead}}.
\newblock In {\em 2015 {{IEEE Symposium}} on {{Security}} and {{Privacy}}},
  pages 866--879.

\bibitem{CHERIHybridCapabilitySystem2015}
Robert~N.M. Watson, Jonathan Woodruff, Peter~G. Neumann, Simon~W. Moore,
  Jonathan Anderson, David Chisnall, Nirav Dave, Brooks Davis, Khilan Gudka,
  Ben Laurie, Steven~J. Murdoch, Robert Norton, Michael Roe, Stacey Son, and
  Munraj Vadera.
\newblock {{CHERI}}: {{A Hybrid Capability-System Architecture}} for {{Scalable
  Software Compartmentalization}}.
\newblock In {\em 2015 {{IEEE Symposium}} on {{Security}} and {{Privacy}}},
  pages 20--37.

\bibitem{SANRAZORReducingRedundant2021}
Jiang Zhang, Shuai Wang, Manuel Rigger, Pinjia He, and Zhendong Su.
\newblock $\{$SANRAZOR$\}$: Reducing redundant sanitizer checks in
  $\{$C/C++$\}$ programs.
\newblock In {\em 15th USENIX Symposium on Operating Systems Design and
  Implementation (OSDI 21)}, pages 479--494, 2021.

\bibitem{DebloatingAddressSanitizer2022}
Yuchen Zhang, Chengbin Pang, Georgios Portokalidis, Nikos Triandopoulos, and
  Jun Xu.
\newblock Debloating address sanitizer.
\newblock In {\em 31st USENIX Security Symposium (USENIX Security 22)}, pages
  4345--4363, 2022.

\bibitem{sactor}
Tianyang Zhou, Haowen Lin, Somesh Jha, Mihai Christodorescu, Kirill Levchenko,
  and Varun Chandrasekaran.
\newblock Llm-driven multi-step translation from {C} to rust using static
  analysis.
\newblock {\em CoRR}, abs/2503.12511, 2025.

\end{thebibliography}

\cleardoublepage
\appendix

\onecolumn
\section{Full Evaluation Results}
\subsection{Check Elision}

\pgfplotstableread[col sep=comma]{data/check_elision_stats.csv}\elisionstats
\pgfplotstableset{
    check_elision_style/.style={
      columns/benchmark/.style={
        column name={Benchmark},
        string type,
        column type={l},
      },
      columns/{safeffi-tim-callgraph: Number of HWASAN Checks}/.style={   
        column name={\# Sanitizer Checks},
        column type={P{1.5cm}|},
        int detect,
        1000 sep={\,},
      },
      columns/{safeffi-tim-callgraph: Elided Checks Percent}/.style={
        column name={Elided Checks},
        column type={P{1.5cm}},
        fixed zerofill,
        precision=2,
        postproc cell content/.append style={
          /pgfplots/table/@cell content/.add={}{\%}
        },
      },
      columns/{safeffi-tim-callgraph: Added Cast Checks}/.style={
        column name={Added Cast Checks},
        column type={P{1.5cm}},
        int detect,
      },
      columns/{safeffi-tim-callgraph: Added Stack Checks}/.style={
        column name={Added Stack Checks},
        column type={P{1.5cm}},
        int detect,
      },
      columns/{safeffi-none: Total Savings}/.style={
        column name={Savings w/o Heap Checks},
        column type={P{1.5cm}|},
        fixed zerofill,
        precision=2,
        postproc cell content/.append style={
          /pgfplots/table/@cell content/.add={}{\%}
        },
      },
      columns/{safeffi-tim-callgraph: Added Heap Checks}/.style={
        column name={Added Heap Checks},
        column type={P{1.5cm}},
        int detect,
      },
      columns/{safeffi-tim-callgraph: Total Savings}/.style={
        column name={Savings w/ Heap Checks},
        column type={P{1.5cm}},
        fixed zerofill,
        precision=2,
        postproc cell content/.append style={
          /pgfplots/table/@cell content/.add={}{\%}
        },
      },
      columns/rustsan: Elided Checks Percent/.style={
        column name={\multicolumn{1}{c}{Savings}},
        column type={||P{1.5cm}}, %
        fixed zerofill,
        precision=2,
        postproc cell content/.append style={
          /pgfplots/table/@cell content/.add={}{\%}
        },
      },
      every head row/.style={
        before row={
          \toprule
          & & \multicolumn{6}{c}{\textbf{SafeFFI}} & \multicolumn{1}{c}{\textbf{RustSan}} \\
          \cmidrule(lr){3-8} %
          \cmidrule(lr){9-9} %
        },
        after row=\midrule
      },
      every last row/.style={after row=\bottomrule},
    }
}

\CommentCameraReady{TODO: vertical line before RustSan show added checks in percent too?}
\begin{table}[!htb]
\label{tab:full_check-elision-stats}
\centering
\scriptsize
\pgfplotstabletypeset[
    check_elision_style,
    columns={
        benchmark,
        {safeffi-tim-callgraph: Number of HWASAN Checks},
        {safeffi-tim-callgraph: Elided Checks Percent},
        {safeffi-tim-callgraph: Added Cast Checks},
        {safeffi-tim-callgraph: Added Stack Checks},
        {safeffi-none: Total Savings},
        {safeffi-tim-callgraph: Added Heap Checks},
        {safeffi-tim-callgraph: Total Savings},
        rustsan: Elided Checks Percent},
]\elisionstats
\end{table}

\subsection{Compile-time Overhead}
\begin{table}[!htb]
\label{fig:full_compile-time-performance}
\centering
\scriptsize
\pgfplotstabletypeset[
  compilation_style,
  columns={
    benchmark,
    baseline_compile_time_s,
    baseline_memory_mb,
    hwasan_compile_time_overhead,
    hwasan_memory_overhead,
    safeffi-none_compile_time_overhead,
    safeffi-none_memory_overhead,
    safeffi-tim-callgraph_compile_time_overhead,
    safeffi-tim-callgraph_memory_overhead,
    rustsan_compile_time_overhead,
    rustsan_memory_overhead
  },
]\compilestats
\end{table}

\subsection{Run-time Overhead}
\begin{table}[!htb]
\label{fig:full_runtime-stats}
\centering
\scriptsize
\pgfplotstabletypeset[
  runtime_style,
  columns={
    Competitor Benchmarks,
    baseline_runtime(ns),
    hwasan_overhead,
    safeffi-none_overhead,
    safeffi-tim-callgraph_overhead
  },
]\runtimestats
\end{table}

\section{ERASan Issue}
\label{appendix:erasan-issue}
\centering
\includegraphics[width=0.4\linewidth, keepaspectratio]{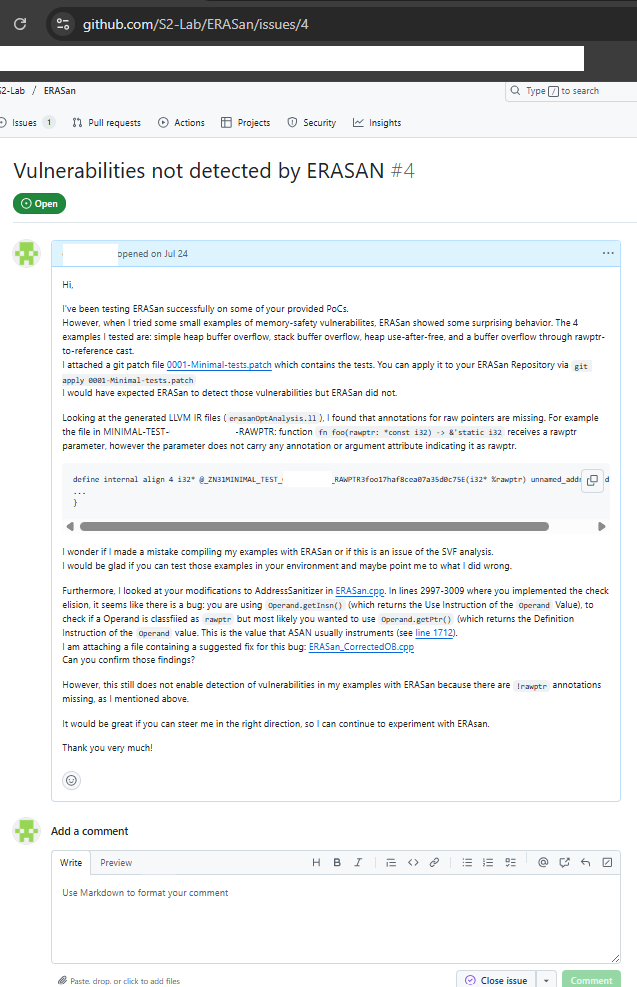}

\end{document}